\documentclass[12pt,preprint,a4paper]{article}

\usepackage{amsmath}
\usepackage{amssymb}
\usepackage{mathrsfs}
\usepackage{graphicx,subcaption}
\usepackage{dsfont}
\usepackage{cite} %for nice listed citations
\usepackage[colorlinks=true,linkcolor=black,citecolor=black,urlcolor=black]{hyperref}
\usepackage{enumerate}
\usepackage{epstopdf}

\numberwithin{equation}{section}

\def\eq#1 { \begin{equation} #1 \end{equation} }

\def\cM{\mathcal{M}}

\def\cM{\mathcal{M}}

\def\sl2r{SL(2,\mathbb{R})}

\def\ff{\gamma}
\def\aa{a}

\def\nc{{\mathrm{nc}}}

\newcommand{\lsim}{\mathrel{\hbox{\rlap{\lower.55ex \hbox{$\sim$}} \kern-.3em \raise.4ex \hbox{$<$}}}}
\newcommand{\gsim}{\mathrel{\hbox{\rlap{\lower.55ex \hbox{$\sim$}} \kern-.3em \raise.4ex \hbox{$>$}}}}

 \newcommand{\be}{\begin{equation}}
\newcommand{\ee}{\end{equation}}

\usepackage{bm}

\makeatletter
\newcommand{\strong}[1]{\@strong{#1}}
\newcommand{\@@strong}[1]{\textbf{\let\@strong\@@@strong#1}}
\newcommand{\@@@strong}[1]{\textnormal{\let\@strong\@@strong#1}}
\let\@strong\@@strong
\makeatother

\begin{document}
%%%%%%%%%%%%%%%%%%%%%%%%%%%%%%%%%%%%%%%%%%%%%%%%%%%%%%%%%%%%%%%%%

\title{\begin{flushright}\vspace{-1in}
       \mbox{\normalsize  EFI-15-13}
       \end{flushright}
       \vskip 20pt
Fields and fluids on curved non-relativistic spacetimes}

\date{\today}

\author{
Michael Geracie
\thanks{\href{mailto:mgeracie@uchicago.edu}         {mgeracie@uchicago.edu}} ,
Kartik Prabhu
\thanks{\href{mailto:kartikp@uchicago.edu}         {kartikp@uchicago.edu}}  and 
   Matthew M. Roberts
   \thanks{\href{mailto:matthewroberts@uchicago.edu}
     {matthewroberts@uchicago.edu}}~
      \\ \\
   {\it Kadanoff Center for Theoretical Physics,}\\
   {\it Enrico Fermi Institute and Department of Physics}\\
   {\it   University of Chicago, Chicago, IL 60637 USA}
} 

\maketitle

\begin{abstract}
We consider non-relativistic curved geometries and argue that the background structure should be generalized  from that considered in previous works. In this approach the derivative operator is defined by a Galilean spin connection valued in the Lie algebra of the Galilean group. This includes the usual spin connection plus an additional ``boost connection'' which parameterizes the freedom in the derivative operator not fixed by torsion or metric compatibility.  As an example we write down the most general theory of dissipative fluids consistent with the second law in curved non-relativistic geometries and find significant differences in the allowed transport coefficients from those found previously. Kubo formulas for all response coefficients are presented. Our approach also immediately generalizes to systems with independent mass and charge currents as would arise in multicomponent fluids. Along the way we also discuss how to write general locally Galilean invariant non-relativistic actions for multiple particle species at any order in derivatives. A detailed review of the geometry and its relation to non-relativistic limits may be found in a companion paper.
\end{abstract}

\newpage
\tableofcontents

%%%%%%%%%%%%%%%%%%%%%%%%%%%%%%%%%%%%%%%%%%%%%%%%%%%%%%%%%%%%%%%%%
%%%%%%%%%%%%%%%%%%%%%%%%%%%%%%%%%%%%%%%%%%%%%%%%%%%%%%%%%%%%%%%%%
\section{Introduction}\label{sec:intro}
%%%%%%%%%%%%%%%%%%%%%%%%%%%%%%%%%%%%%%%%%%%%%%%%%%%%%%%%%%%%%%%%%

Newton-Cartan geometry was first developed shortly following the inception of general relativity in an attempt to phrase non-relativistic physics in a manner that respects general coordinate invariance \cite{Cartan:1923zea,Cartan:1924yea} and later developed further in \cite{DH,Trautman:1965ul,Kunzle:1972fv,Duval:1976ht,Kuchar-Sch,DK,Duval:1984cj,Banerjee:2014pya,Banerjee:2014nja,Banerjee:2015tga}. Initial studies concerned themselves with only the spacetime structure; however, recent applications in condensed matter physics have focused on covariantly coupling matter to this background. Perfect fluids in non-relativistic backgrounds were first studied in \cite{Carter:1993aq,Carter:2003im} while applications to the fractional quantum Hall effect have proven exceptionally rich \cite{Hoyos:2011ez,Son:2013,Abanov:2014ula,Gromov:2014vla}, ranging from gravitational anomalies to energy transport. Newton-Cartan geometry has also naturally arisen in the study of non-relativistic holographic systems, where the boundary theory realizes a ``twistless-torsionful'' Newton-Cartan geometry \cite{Christensen:2013lma, Christensen:2013rfa, Bergshoeff:2014uea, Hartong:2014pma, Hartong:2014oma, Hartong:2015wxa, Hartong:2015zia}.

One of the principle benefits of the geometric approach is that it furnishes a collection of background data consistent with Galilean invariance that may be perturbed to covariantly define currents and study linear response. These include a ``clock'' one form $n_\mu$ defining a preferred notion of spatial vectors and elapsed time as well as a ``spatial inverse metric'' $h^{\mu \nu}$ satisfying
\begin{align}
	h^{\mu \nu} n_\nu = 0.
\end{align}
In the presence of a conserved particle current, one also has the option to couple to a background $U(1)$ connection $A_\mu$. In the case of a single Schr\"odinger field we then have
\begin{align}
	S = \int d^3 x \sqrt{h} \left( \frac{i}{2} \psi^\dagger \overset{\leftrightarrow} D_0 \psi - \frac{h^{ij}}{2m} D_i \psi^\dagger D_j \psi \right)
\end{align}
where $D_\mu = \partial_\mu - i A_\mu$, $h^{ij}$ are the spatial components of the metric $h^{\mu \nu}$ and we have taken $n_\mu = \begin{pmatrix} 1 & 0 \end{pmatrix}$ for simplicity. 

It was noted in \cite{Son:2005rv} that for this action to be invariant under arbitrary space and time dependent changes of coordinates, the vector potential must carry the anomalous transformation law
\begin{align}
	\delta A_0 = - \xi^\lambda \partial_\lambda A_0 - A_\lambda \dot \xi^\lambda ,
	&&\delta A_i = - \xi^\lambda \partial_\lambda A_i - A_\lambda \partial_i \xi^\lambda - m h_{ij} \dot \xi^j
\end{align}
where $\xi^\mu$ is an infinitesimal diffeomorphism. Though simple in form, the final term is rather curious as $A_\mu$ does not transform as a one form. The origin of this term was properly isolated in \cite{Jensen:2014aia}, identifying an implicit change in rest frame in the transformation above. The vector potential is then a true one form under diffeomorphisms,
\begin{align}
	\delta A_\mu = - \xi^\lambda \partial_\lambda A_\mu - A_\lambda \partial_\mu \xi^\lambda
\end{align}
while under a Galilean boost in the manner
\begin{align}\label{A transf}
	A_\mu \rightarrow A_\mu +  k_\mu - \frac{1}{2} n_\mu k^2 ,
\end{align}
where $k_\mu$ is a spatial vector representing the boost velocity.
 
The background data is then the triple $(n, h, A)$ up to a boost transformation given by (\ref{A transf}). In this paper, we demonstrate that this is not the largest collection of data consistent with Galilean invariance and augment the background accordingly. Our list of background fields includes
\begin{equation}\label{our data}
	e^A_\mu , \hspace{30pt}
	\omega_\mu^A{}_B,\hspace{30pt}
	\aa_\mu,\hspace{30pt}
	A_\mu.
\end{equation}
Here $e^A_\mu$ is a veilbein transforming as a Galilean vector under local boosts and rotations and is equivalent in content to the clock form and spatial metric considered above. $\omega_\mu^A{}_B$ is a connection one-form valued in the Lie algebra of the Galilean group. It contains the usual spin connection $\omega_\mu^{ab} = \omega_\mu^{[ab]}$ but also a  ``boost connection'' $ \varpi^a = \omega^a{}_0$ which transforms under a choice of reference frame.  We have also separated out the single gauge field appearing in previous treatments to two gauge fields $\aa$ and $A$ so that we may describe independent mass (or particle number) and charged currents. It is then the mass gauge field $\aa$ that transforms in the manner (\ref{A transf}), while the electromagnetic gauge field $A$ is boost invariant.

This treatment also resolves a lingering question of \cite{Jensen:2014aia}. With the data $(n, h, A)$ at hand, one may define a connection
\begin{align}\label{jensen connection}
	{\Gamma^\lambda}_{\mu \nu} = v^\lambda \partial_{\mu} n_{\nu} + \frac{1}{2} h^{\lambda \rho} \left( \partial_\mu h_{\nu \rho} + \partial_\nu h_{\mu \rho} - \partial_\rho h_{\mu \nu} \right) + n_{(\mu} F_{\nu ) \rho}h^{\rho \lambda},
\end{align}
where $F = d A$ is the field strength of $A$. Here $v^\mu$ is a vector field parameterizing the choice of rest frame and $h_{\mu \nu}$ is defined by (\ref{lowered metric}). This connection is boost invariant and so defines a sensible, frame independent geometry in the absence of torsion, but fails to do so generically. However, the Galilean spin-connection $\omega^A{}_B$ defines an invariant connection $\nabla$ on all backgrounds.
In a companion paper \cite{GPR_geometry}, we consider the suite of possible constraints that may be placed on (\ref{our data}) in a boost invariant manner. In particular, the connection (\ref{jensen connection}) may be obtained by the identification
\begin{align}
	\varpi^a \wedge e_a =  d \aa
\end{align}
but we find that this is only possible when the torsion vanishes.

One might reasonably hope that this issue may be avoided since torsion vanishes on a large class of physically relevant backgrounds\footnote{Torsion can be relevant in the study of elasticity \cite{landau1986theory} and lattice defects \cite{Hughes:2012vg}.}. However, studying energy transport requires the introduction of a Luttinger potential $\Phi$ \cite{Luttinger:1964zz}, which arises in the spacetime approach as temporal torsion \cite{Geracie:2014nka, Gromov:2014vla}. Thus if we hope to use Newton-Cartan geometry to study energy currents, we must know that our derivative operator respects Galilean invariance in such a background.
To this end, we shall consider non-relativistic fluids coupled to (\ref{our data}) and perform an entropy current analysis and find results that do not agree with those of \cite{Jensen:2014ama}. 

We begin in section \ref{sec:gal} by reviewing the basics of Newton-Cartan geometry, boost transformations and representations of the Galilean group that will be needed for our later analysis. Section \ref{sec:geo} then introduces the mass gauge field $\aa$ and boost connection $\varpi^a$. Before continuing on to fluids, we present in section \ref{sec:actions} a brief detour through Galilean invariant actions and show how to write down actions for massive non-relativistic fields at any order in derivatives. The approach is seen to be equivalent to null reduction, but is intrinsic to the Newton-Cartan spacetime.

Finally we develop first order fluid dynamics in our approach, beginning with a manifestly boost covariant presentation of the full set of diffeomorphism Ward identities in section \ref{sec:ward}. Section \ref{sec:fluids} then presents the fluid equations of motion and performs the entropy current analysis. For multicomponent fluids, the results are summarized as follows (all coefficients are arbitrary functions of the temperature $T$, charge chemical potential $\mu_Q$ and mass chemical potential $\mu_M$ unless stated otherwise).
The most general set of first order transport coefficients in the parity even sector includes four sign semi-definite functions: a bulk viscosity, shear viscosity, conductivity and thermal conductivity
\begin{align}
	\zeta \geq 0,
	&&\eta \geq 0,
	&&\sigma_E \geq 0 ,
	&&\kappa_T \leq 0.
\end{align}
as well as a thermo-electric coefficient
\begin{align}
	\sigma_T .
\end{align}
The parity odd sector contains a Hall viscosity, Hall conductivity, thermal Hall conductivity, thermo-electric Hall coefficient, magnetization and energy magnetization
\begin{align}
	\tilde \eta,
	&&\tilde \sigma_E,
	&&\tilde \kappa_T,
	&&\tilde \sigma_T ,
	&& \tilde m,
	&& \tilde m_\epsilon.
\end{align}
The magnetization determines the magnetic field induced pressure via the coefficient
\begin{align}
	&\tilde f_B =T^2 \partial_\epsilon p  \partial_T \left(  \frac{\tilde m}{T} \right) + \partial_q p \partial_Q  \left(  \frac{\tilde m}{T} \right) + \partial_\rho p \partial_M  \left(  \frac{\tilde m}{T} \right) ,
\end{align}
where $p ( \epsilon , q , \rho)$ is the pressure as a function of energy, charge and mass density. Kubo formulas for these coefficients are then presented.

In flat backgrounds with no Luttinger potential the constitutive relations are
\begin{align}
	&\rho^0 = \rho, \qquad \qquad \rho^i = \rho u^i
	\qquad \qquad j^0 = q , 
	\qquad \qquad  \varepsilon_\nc^0 = \frac{1}{2} \rho u^2 + \epsilon ,\nonumber \\
	 &j^i = q u^i + \sigma_E (  E^i + B \epsilon^{ij} u_j - T \partial^i \nu_Q ) + \tilde \sigma_E \epsilon^{ij} (  E_j + B \epsilon_{jk} u^k - T \partial_j \nu_Q ) \nonumber \\
	 &\qquad + \sigma_T  \partial^i T + \tilde \sigma_T  \epsilon^{ij} \partial_j T + \epsilon^{ij} \partial_j \tilde m , \nonumber \\
	 & \varepsilon_\nc^i = \left( \frac{1}{2} \rho u^2 + \epsilon + p - \zeta \theta - \tilde f_B B \right) u^i - \eta \sigma^{ij} u_j - \tilde \eta \tilde \sigma^{ij} u_j  \nonumber \\
	 &\qquad + T \sigma_T (  E^i + B \epsilon^{ij} u_j - T \partial^i \nu_Q ) - T \tilde \sigma_T \epsilon^{ij} ( E_j + B \epsilon_{jk} u^k - T \partial_j \nu_Q ) \nonumber \\
	 &\qquad + \kappa_T  \partial^i  T + \tilde \kappa_T  \epsilon^{ij} \partial_j T -\tilde m \epsilon^{ij} (E_j + B \epsilon_{jk} u^k ) +  \epsilon^{ij} \partial_j  \tilde m_\epsilon  , \nonumber \\
	 & T_\nc^{ij} = \rho u^i u^j + ( p - \zeta \theta - \tilde f_B B ) g^{ij} - \eta \sigma^{ij} - \tilde \eta \tilde \sigma^{ij} .
\end{align}
where $\rho^\mu$ is the mass current, $j^\mu$ the  charge current, $ \varepsilon_\nc^\mu$ the energy current  and $ T^{ij}_\nc$ the spatial stress and the fluid shear $\sigma^{ij}$ and expansion $\theta$ are defined by
\begin{align}
	&\sigma_{ij} =\partial_i u_j + \partial_i u_i - \delta_{ij} \theta ,
	&&\theta = \partial_i u^i  .
\end{align}
The epsilon symbols are chosen with sign convention $\epsilon^{12} = \epsilon_{12} = 1$.
In summarizing these results we have made the choice of fluid frame (\ref{frame}).

{\bf Note added:} It was noted in \cite{Banerjee:2015hra} that our result misses a vorticity induced pressure coefficient due to a technical error. The correction may be found in (C.3) of their work.

\section{Galilean Symmetry}\label{sec:gal}

The salient feature of non-relativistic physics is that of Galilean relativity, which, in its most familiar form, asserts that the laws of physics do not depend on a choice of inertial reference frame. These frames are related by the Galilean transformations 
\begin{align}\label{galilean}
	&t \rightarrow t
	&&x^i \rightarrow {\Theta^i}_j x^j - k^i t .
\end{align}
Here ${\Theta^i}_j \in SO(d)$ determines the relative orientation and $k^i \in \mathbb R^d$ the relative velocity of inertial observers adopting coordinates $(t ,x^i)$ and $(t',x'^i)$.

In curved space one in general loses a notion of inertial frames and preferred coordinate systems. Rather, one can define only coframes
\begin{align}
	e_\mu^A,
	\qquad \qquad
	A = 0, 1, \cdots d
\end{align}
which form a local basis of 1-forms on the spacetime manifold $\mathcal M$, which we take to be $d+1$ dimensional. The proper coordinate invariant statement of (\ref{galilean}) is then that these coframes transform as
\begin{align}\label{fundamental}
	\begin{pmatrix}
		e^0 \\
		e^a
	\end{pmatrix}
	\rightarrow
	\begin{pmatrix}
		1 & 0 \\
		-k^a & {\Theta^a}_b
	\end{pmatrix}
	\begin{pmatrix}
		e^0 \\
		e^b
	\end{pmatrix}.
\end{align}
Here spatial indices running over the values $1,\dots,d$ are denoted by lower case Latin letters $a,b,\dots$ to distinguish them from spacetime indices $A,B,\dots$. Greek letters $\mu , \nu , \dots$ will represent coordinate indices while $i,j,\dots$ will be their spatial components. Of course, if the veilbein can be chosen to correspond to a global coordinate basis $e^A = d x^A$, we retrieve (\ref{galilean}).
Given a coframe $e^A_\mu$ we may of course define a set of frame vectors $e_A^\mu$ such that
\begin{align}
	e^A_\mu e^\mu_B = {\delta^A}_B
	&&
	e^A_\mu e^\nu_A = {\delta^\nu}_\mu
\end{align}
which transforms via the inverse of (\ref{fundamental}),
\begin{align}\label{anti-fundamental}
	\begin{pmatrix}
		e_0 &e_a
	\end{pmatrix}
	\rightarrow
	\begin{pmatrix}
		e_0 &
		e_b
	\end{pmatrix}
	\begin{pmatrix}
		1 & 0 \\
		(\Theta^{-1})^b{}_c k^c & (\Theta^{-1})^b{}_a
	\end{pmatrix} .
\end{align}

It's easily seen from (\ref{fundamental}) and (\ref{anti-fundamental}) that these transformations preserve the spacetime tensors
\begin{align}
	&n_\mu \equiv e^0_\mu
	&&h^{\mu \nu} \equiv e^\mu_a e^\nu_b \delta^{ab} .
\end{align}
The natural geometric setting for non-relativistic physics thus involves a positive semi-definite symmetric $(2,0)$ tensor $h^{\mu \nu}$ and 1-form $n_\mu$. These tensors are related insofar as $n_\mu$ spans the single degeneracy direction of $h^{\mu \nu}$
\begin{align}
	h^{\mu \nu} n_\nu = 0.
\end{align}

These two fields are usually taken as the starting point in defining a Newton-Cartan geometry, though we have chosen rather to go through a veilbein formalism since this will prove most convenient for our later analysis. They have clear physical interpretations. The ``clock-form'' $n_\mu$ defines a preferred notion of spatial direction at each point as well as an arrow of time: vector fields $t^\mu$ such that
\begin{align}
	n_\mu t^\mu > 0
\end{align}
being future directed. Any curve $\gamma$ also inherits a notion of elapsed time
\begin{align}\label{elapsed time}
	\Delta T = \int_\gamma n 
\end{align}
while $h^{\mu \nu}$ serves as a spatial ``inverse metric''\footnote{The terminology can be deceptive. Since $h^{\mu \nu}$ is degenerate it is neither invertible nor a metric. However, since it contains precisely enough data to define a unique Riemannian metric on spatial slices this terminology should not provoke undue confusion.}.

The clock form defines a pointwise notion of spatial directions via vectors $w^\mu$ such that $n_\mu w^\mu = 0$. However, the Frobenius theorem tells us this notion may be integrated to a local codimension-1 hypersurface if and only if $n \wedge dn = 0$. For our non-relativistic spacetime to carry an (at least local) notion of simultaneity, we must then demand that this holds everywhere. There are far more pressing reasons to take $n \wedge dn = 0$ however, for a theorem due to Carath\'eodory \cite{Frankel} ensures the existence of closed timelike curves passing through any point violating this condition. To be precise, if $n \wedge dn \neq 0$ at the point $p \in \cM$, then there exists a neighborhood of $p$ in which any two points may be connected by a future directed timelike curve\footnote{The cited theorem actually concerns null curves. However, it is easy to extend this result to timelike curves by adding a very small future directed component.}. We thus refer to spacetimes with $n \wedge dn = 0$ everywhere as causal and will only consider such spaces throughout.

In causal spacetimes, one may always choose coordinates such that $n$ has no spatial components, and so the metric and clock form take the form
\begin{align}\label{coords}
	n_\mu =
	\begin{pmatrix}
		e^{-\Phi} & 0 \\
	\end{pmatrix},
	&&
	h^{\mu \nu} =
	\begin{pmatrix}
		0 & 0 \\
		0 & h^{ij}
	\end{pmatrix}
\end{align}
where $h^{ij}$ is everywhere a metric of Riemannian signature.

\subsection{The Galilean Group}

A few words on Galilean representations will prove helpful in what follows.
The matrices appearing in (\ref{fundamental}) form a group under matrix multiplication called the Galilean group $Gal(d)$. Throughout we shall refer to this defining representation as the vector representation and denote matrices in this representation as $\Lambda^A{}_B$. Coframes then transform in the vector representation of $Gal(d)$ and frames in the dual
\begin{align}\label{galilean2}
	&e^A \rightarrow {\Lambda^A}_B e^B, 
	&& e_A \rightarrow e_B (\Lambda^{-1})^B{}_A .
\end{align}

There is an equivalent definition of $Gal(d)$ in terms of invariant tensor data. One may check that the matrices $\Lambda^A{}_B$ are the unique ones that leave unchanged
\begin{align}\label{fundamental invariants}
	&n_A = 
	\begin{pmatrix}
		1 & 0 
	\end{pmatrix},
	&&h^{AB} = 
	\begin{pmatrix}
		0 & 0 \\
		0 & \delta^{ab}
	\end{pmatrix} .
\end{align}
This is the reason that a NC geometry contains precisely an invariant 1-form and degenerate spatial metric from a representation theoretic point of view, for the tensors
\begin{align}
	n_\mu = n_A e^A_\mu,
	&&h^{\mu \nu} = e^\mu_A e^\nu_B h^{AB}
\end{align}
are then the primitive Galilean invariants that may be formed from the veilbein.

Since $Gal(d)$ is a subgroup of $SL(d+1,\mathbb R )$, the epsilon symbols $\epsilon_{A_0 \cdots A_d}$ and $\epsilon^{A_0 \cdots A_d}$ with $\epsilon_{01\cdots d} = \epsilon^{01\cdots d} = 1$ are also invariant tensors. We may use them to define a spacetime volume element
\begin{align}
	\varepsilon_{\mu_0 \cdots \mu_d} = \epsilon_{A_0 \cdots A_d}e^{A_0}_{\mu_0} \cdots e^{A_d}_{\mu_d}
\end{align}
as well as a ``raised volume element"
\begin{align}
	\varepsilon^{\mu_0 \cdots \mu_d} = \epsilon^{A_0 \cdots A_d}e_{A_0}^{\mu_0} \cdots e_{A_d}^{\mu_d} .
\end{align}
We caution the reader however that $\varepsilon^{\mu_0 \cdots \mu_d}$ is not obtained from $\varepsilon_{\mu_0 \cdots \mu_d}$ by raising indices since the metric $h^{\mu \nu}$ is degenerate. These are both annihilated by the derivative operator that we shall define and in coordinates where (\ref{coords}) holds,
\begin{align}
	\varepsilon_{01\cdots d} = \sqrt{|h|} e^{- \Phi},
	&&
	\varepsilon^{01 \cdots d} = \frac{1}{\sqrt{|h|} e^{- \Phi}} .
\end{align}
We can also naturally contruct an invariant spatial volume element (with raised indices only) $\varepsilon^{\mu_1\ldots\mu_d} = \varepsilon^{\mu_1\ldots\mu_d\rho} n_\rho$.

\subsection{The Extended Representation}

Another representation of $Gal(d)$ will prove remarkably useful when we turn to writing Galilean invariant actions (section \ref{sec:actions}) as well as in presenting manifestly covariant fluid dynamics. Explicitly, it is of the form
\begin{align}\label{extended}
	{\Lambda^I}_J =
	\begin{pmatrix}
		1 & 0 & 0 \\
		- k^a & {\Theta^a}_b & 0 \\
		- \frac{1}{2} k^2 & k_c {\Theta^c}_b & 1 
	\end{pmatrix}
\end{align}
where $k^2 = k^a k_a = \delta_{ab} k^a k^b$. We shall refer to this representation as the extended representation. It is $d+2$ dimensional and has it's own set of defining invariant tensors
\begin{align}\label{extended invariants}
	&n_I = 
	\begin{pmatrix}
		1 & 0 & 0 \\
	\end{pmatrix},
	&&g^{IJ}
	=
	\begin{pmatrix}
		0 & 0 & 1 \\
		0 & \delta^{ab} & 0 \\
		1 & 0 & 0
	\end{pmatrix} .
\end{align}
Here the indices $I,J,\dots$ run from $0$ to $d+1$ and will always denote an object in the extended/dual-extended representation. The convention used in (\ref{extended}) is that the first row/column indicates the time component $0$, the second indicates the full set of spatial components $a$ and the final one the $(d+1)^\mathrm{th}$ component. 

Although less familiar than the vector, the extended representation is often easier to work with since it admits the Lorentzian metric $g^{IJ}$. Whenever working with this representation we will freely raise and lower indices with $g^{IJ}$ in the typical manner. Furthermore, any object in the extended representation may be projected to a Galilean vector via application of
\begin{align}\label{projector}
	\Pi^A{}_I = 
	\begin{pmatrix}
		\delta^A{}_B & 0
	\end{pmatrix}
\end{align} 
which is also invariant under the action of $Gal(d)$.

\subsection{Timelike Vector Fields and Milne Symmetry}
We thus see that non-relativistic geometries are naturally defined by a preferred clock-form $n_\mu$ and a metric $h^{\mu \nu}$ that annihilates it. However, one thing that it cannot include is a preferred timelike vector field $v^\mu$, for any such field is equivalent to a preferred notion of rest\footnote{Of course, this may be sensible in the presence of additional fields; for example, a background magnetic field or lattice establishes a preferred notion of rest with respect to which all velocities may be measured. However, the spacetime structure itself can make no such distinction.}. It is nonetheless often convenient to introduce such a $v^\mu$ for use in explicit formulas (for instance, in computing the Christoffel symbols) and is essential for writing time derivatives (and thus any dynamical equation - see section {\ref{sec:actions}}). We will always normalize $v^\mu$ in the sense
\begin{align}\label{v norm}
	v^\mu n_\mu = 1 .
\end{align}

Given such a $v^\mu$ we may then define a transverse projector 
\begin{align}
	{P^\mu}_\nu = {\delta^\mu}_\nu - v^\mu n_\nu 
\end{align}
whose upper index is $n_\mu$ orthogonal and whose lower index is $v^\mu$ orthogonal as well as a ``lowered spatial metric'' $h_{\mu \nu}$ that ``inverts'' the degenerate $h^{\mu \nu}$ to this projector
\begin{align}\label{lowered metric}
	h_{\mu \nu} v^\nu = 0,
	&&h^{\mu \lambda} h_{\lambda \nu} = {P^\mu}_\nu .
\end{align}
In coordinates (\ref{coords}) we have
\begin{align}
	v^\mu =
	\begin{pmatrix}
		e^\Phi \\
		v^i
	\end{pmatrix},
	&&P^\mu{}_\nu =
	\begin{pmatrix}
		0 & 0 \\
		- e^{-\Phi} v^i & \delta^i{}_j
	\end{pmatrix},
	&&h_{\mu \nu} = 
	\begin{pmatrix}
		e^{-2 \Phi} v^2 & - e^{-\Phi} v_j \\
		- e^{-\Phi} v_i & h_{ij}
	\end{pmatrix}.
\end{align}

Since $v^\mu$ has no physical significance, there must be a symmetry under shifts by a spatial vector $k^\mu$ so that (\ref{v norm}) is preserved
\begin{align}\label{milne}
	v^\mu \rightarrow v^\mu + k^\mu 
	&& n_\mu k^\mu = 0
\end{align}
This symmetry is often referred to in the literature as a ``Milne boost'' \cite{Duval:2009vt}, and the projector and lowered metric also transform under this redefinition
\begin{align}\label{milne2}
	{P^\mu}_\nu \rightarrow {P^\mu}_\nu - k^\mu n_\nu ,
	&&h_{\mu \nu} \rightarrow h_{\mu \nu} - n_\mu k_\nu - k_\mu n_\nu + k^2 n_\mu n_\nu,
\end{align} 
where $k_\mu = h_{\mu \nu} k^\nu$ and $k^2 = h_{\mu \nu} k^\mu k^\nu$.
Note that if $n = e^{-\Phi} dt$ then the pullback of $h_{\mu\nu}$ defines an invariant non-degenerate metric on constant $t$ hypersurfaces. If we adopt a veilbein formalism for Newton-Cartan geometry, there is a natural timelike vector field in the form of $e^\mu_0$. Since $e^\mu_0$ transforms under local Galilean boosts in the same manner as (\ref{milne}) it is natural to identify
\begin{align}
	v^\mu = e^\mu_0
\end{align}
and so Milne boosts and local Galilean boosts are also identified. In the veilbein formalism we then also have
\begin{align}
	{P^\mu}_\nu = e^\mu_a e^a_\nu,
	&&h_{\mu \nu} = e^a_\mu e^b_\nu \delta_{ab}
\end{align}
and the transformations (\ref{milne2}) again follow precisely from (\ref{fundamental}) and (\ref{anti-fundamental}).

\section{Non-relativistic Geometry}\label{sec:geo}

To complete our description of non-relativistic spacetimes we also require a notion of parallel transport. This is specified via a connection $\nabla$ that preserves the above data
\begin{align}\label{metric compatibility}
	\nabla_\mu n_\nu = 0 ,
	&&\nabla_\lambda h^{\mu \nu} = 0 .
\end{align}
In the Lorentzian case, metric compatibility completely determines the connection (up to torsion). Unfortunately, as is well known, (\ref{metric compatibility}) does not do so in a Newton-Cartan geometry and there are many distinct connections (even if we require vanishing torsion) all satisfying these conditions.

%%%%%%%%%%%%%%%%%%%%%%%%%%%%%%%%%%%%%%%%%%%%%%%%%%%%%%%%%%%%%%%%%%%%
%%%%%%%%%%End New Stuff
%%%%%%%%%%%%%%%%%%%%%%%%%%%%%%%%%%%%%%%%%%%%%%%%%%%%%%%%%%%%%%%%%%%%

Various authors have dealt with this ambiguity in different ways. In the recent condensed matter literature, this extra freedom has been fixed by introducing additional conditions to further constrain the connection. In \cite{Son:2013} this was done by demanding the curl-freeness of the vector field $v^\mu$
\begin{align}
	h_{\lambda [ \mu} \nabla_{\nu ]} v^\lambda = 0 .
\end{align}
The associated Christoffels are
\begin{align}
	{\Gamma^\lambda}_{\mu \nu} = v^\lambda \partial_{(\mu} n_{\nu)} + \frac{1}{2} h^{\lambda \rho} \left( \partial_\mu h_{\nu \rho} + \partial_\nu h_{\mu \rho} - \partial_\rho h_{\mu \nu} \right) ,
\end{align}
where we have assumed torsion-freeness for simplicity.
Of course, since $v^\mu$ and $h_{\mu\nu}$ are not a boost invariant quantities, the connection so defined is not either. 

As mentioned in the discussion below (\ref{jensen connection}), recent works \cite{Jensen:2014aia} have taken the approach that this freedom should be fixed by the field strength $F_{\mu \nu}$ of the gauge field coupling to particle current, asserting rather
\begin{align}\label{restriction}
	h_{\lambda [ \mu} \nabla_{\nu ]} v^\lambda = - \frac{1}{2} F_{\mu \nu} .
\end{align}
giving Christoffels
\begin{align}
	{\Gamma^\lambda}_{\mu \nu} = v^\lambda \partial_{(\mu} n_{\nu)} + \frac{1}{2} h^{\lambda \rho} \left( \partial_\mu h_{\nu \rho} + \partial_\nu h_{\mu \rho} - \partial_\rho h_{\mu \nu} \right) + n_{(\mu} F_{\nu )}{}^\lambda,
\end{align}
where one of the $F_{\mu \nu}$ indices has been raised with a $h^{\mu \nu}$. With the transformation law
\begin{align}\label{A boost}
	A_\mu \rightarrow A_\mu + k_\mu - \frac{1}{2} n_\mu k^2,
\end{align}
this defines a boost invariant connection in torsion-free backgrounds.

However, as discussed in \cite{Jensen:2014aia} and reviewed in section \ref{sec:intro}, this is lost upon the introduction of torsion. The resulting notion of parallel transport then depends on a notion of rest frame and so is inconsistent with the principle of Galilean relativity. In \cite{GPR_geometry} we consider the possible restrictions consistent with boost invariance that one may place on the connection and find limited freedom when the background is arbitrary. We thus take the point of view that the connection should not be restricted when studying energy transport.

\subsection{Newton-Cartan Geometry}\label{subsec:NC_geom}
In our approach then, a Newton-Cartan geometry is a $d+1$-dimensional manifold $\mathcal M$ with a collection $(n,h,\nabla)$ of a 1-form $n_\mu$, a rank-$d$ positive semi-definite metric $h^{\mu \nu}$, and a connection $\nabla$ such that
\begin{align}\label{NC def}
	h^{\mu \nu} n_\nu = 0,
	&&\nabla_\mu n_\nu = 0,
	&&\nabla_\lambda h^{\mu \nu} = 0 .
\end{align}
The connection is then simply extra data above and beyond what appears in the clock-form and spatial metric and is boost invariant by definition.

The additional data may be parameterized in a boost non-invariant way by introducing a $v^\mu$. It's derivative is some tensor
\begin{align}\label{v connection}
	\nabla_\mu v^\nu = \Lambda_\mu{}^\nu
\end{align}
that includes information on the acceleration, shear, expansion and twist of the vector field $v^\mu$. The Christoffel symbols are then
\begin{align}\label{christoffels}
	{\Gamma^\lambda}_{\mu \nu} = v^\lambda \partial_{(\mu} n_{\nu)} + \frac{1}{2} h^{\lambda \rho} \left( \partial_\mu h_{\nu \rho} + \partial_\nu h_{\mu \rho} - \partial_\rho h_{\mu \nu} \right) + \frac{1}{2} \left( T^\lambda{}_{\mu \nu} - T_{\mu \nu}{}^\lambda - T_{\nu \mu}{}^\lambda \right) + n_{(\mu} \Omega_{\nu )}{}^\lambda,
\end{align}
where we have also included possible non-zero torsion $T^\lambda{}_{\mu \nu}$. These Christoffel symbols were recently derived in the formalism of Koszul connections in \cite{BM}. For a veilbein approach see \cite{GPR_geometry}.  Here $\Omega_{\mu \nu} \equiv 2 \Lambda_{[\mu \nu ]}$ involves the vorticity and acceleration of $v^\mu$, which contains the data unfixed by the torsion and (\ref{NC def}). We shall refer to this as the Newton-Coriolis 2-form. One may check that under boosts, $\Omega$ transforms as
\begin{align}\label{coriolis transformation}
	\Omega_{\mu \nu}  \rightarrow \Omega_{\mu \nu}  + 2 \partial_{[\mu} ( k_{\nu ] } - \frac{1}{2} n_{\nu ]}  k^2 ) - ( k_\lambda - \frac{1}{2} n_\lambda k^2 ) T^\lambda{}_{\mu \nu}
\end{align}
so that (\ref{christoffels}) is invariant even in the presence of torsion. In deriving (\ref{coriolis transformation}) we have made liberal use of the identity $\nabla_\lambda h_{\mu \nu} = - 2 \Lambda_{\lambda ( \mu} n_{\nu )}$ which is an immediate consequence of the definitions.

In relativistic physics we work with Lorentzian geometry where the torsion may be chosen independently of the metric, but it is well known that the analogous statement is not true for non-relativistic physics. Anti-symmetrizing the expression $\nabla_\mu n_\nu = 0$ gives the condition
\begin{align}\label{temporal torsion}
	n_\lambda {T^\lambda}_{\mu \nu} = (d n)_{\mu \nu} .
\end{align}
Although we expect $dn=0$ on most physically relevant non-relativistic spacetimes\footnote{$dn =0$ is equivalent to a notion of absolute time. For details we refer to the discussion in \cite{GPR_geometry}.}, it is useful to keep this torsion around at intermediate stages when studying energy transport. For this reason it is important to know that our geometry is sensible in the presence of torsion.

\subsection{A Galilean Spin Connection}
The data contained in a Newton-Cartan geometry can be equivalently stated purely in a veilbein formalism. In this approach, rather than the clock form $n_\mu$ and spatial metric $h^{\mu \nu}$, we have the coframe $e^A_\mu$ transforming as a Galilean vector. The complete data of the connection is then equivalent to a Galilean spin connection $\omega^A{}_B$ defined by
\begin{align}\label{spin connection}
	\nabla_\mu e^A_\nu = - \omega_\mu{}^A{}_B e^B_\nu , \qquad \implies \qquad
	 \nabla_\mu e^\nu_A =  e_B^\nu \omega_\mu{}^B{}_A 
\end{align}
which implies the usual transformation law for connections
\begin{align}
	\omega^A{}_B \rightarrow {\Lambda^A}_C \omega^C{}_D (\Lambda^{-1})^D{}_B + {\Lambda^A}_C d (\Lambda^{-1})^C{}_B 
\end{align}
for an arbitrary local Galilean transformation $\Lambda^A{}_B$. The curvature and torsion are then defined in the usual way
\begin{align}\label{curvTor}
	R^A{}_B = d \omega^A{}_B + \omega^A{}_C \wedge \omega^C{}_B ,
	&&T^A = D e^A = d e^A + \omega^A{}_B \wedge e^B
\end{align}
and are boost covariant.

The compatibility conditions (\ref{metric compatibility}) then imply 
\begin{align}
	0 &=  e^\nu_A \nabla_\mu e^0_\nu = - \omega_\mu{}^0{}_A ,
	&&0 =e^a_\mu e^b_\nu \nabla_\lambda (e_C^\mu  e_D^\nu h^{CD} ) = 2 \omega_\lambda{}^{(ab)} .
\end{align}
That is, they are equivalent to $\omega^A{}_B$ being in the Lie algebra of the Galilean group $gal(d)$
\begin{align}
	\omega^A{}_B =
	\begin{pmatrix}
		0 & 0 \\
		\varpi^a & \omega^a{}_b
	\end{pmatrix}
\end{align}
where $\omega_{\mu ab} = \omega_{\mu [ a b ]}$. The spin part $\omega_\mu{}^{a}{}_b$ of the Galilean connection is familiar. It defines the connection on a spatial slice and may be used to covariantize actions involving fields that transform under local rotations through the covariant derivative
\begin{align}
	\psi \rightarrow e^{- \frac{i}{2} \theta^{ab} J_{ab} }\psi, 
	&&D_\mu = \partial_\mu + \frac{1}{2} \omega_\mu{}^{ab} J_{ab}  .
\end{align}
where $J_{ab}$ are the generators of rotations. 

The boost part $\varpi_\mu^a$ is, however, new. It transforms as a vector under rotations but as a connection under boosts
\begin{align}
	\varpi^a \rightarrow \varpi^a +  D k^a .
\end{align}
where $D k^a = d k^a + \omega^a{}_b k^b$. It is the boost connection that parameterizes the freedom in the metric compatibility conditions and is equivalent to the tensor $\Lambda_\mu{}^\nu$ considered earlier
\begin{align}\label{sup}
	\Lambda_\mu{}^\nu = \nabla_\mu v^\nu = e_a^\nu \varpi^a_\mu .
\end{align}
It is then our point of view that a non-relativistic geometry is specified by the pair $(e^A , \omega^A{}_B)$, which is equivalent  to the treatment of the previous section.

We shall occasionally find it useful to consider the spin-connection valued in the extended representation
\begin{align}
	\omega^I{}_J = 
	\begin{pmatrix}
		0 & 0 & 0 \\
		\varpi^a & \omega^a{}_b & 0 \\
		0 & - \varpi_b & 0 
	\end{pmatrix} .
\end{align}
Note that the $I,J$ indices here indicate a homomorphism of $\omega^A{}_B$ from the defining to the extended representation of the Lie algebra, and not application of any of the invariant tensors considered previously.

\section{Non-Relativistic Actions}\label{sec:actions}
We now have the necessary tools to present non-relativistic actions in a manifestly invariant manner. We begin with the Schr\"odinger action in $d$ dimensional flat space
\begin{align}\label{flat action}
	S = \int d^{d+1} x \left( \frac{i}{2} \psi^\dagger \overset{\leftrightarrow} D_0 \psi - \frac{\delta^{ij}}{2m}(D_i \psi )^\dagger (D_j \psi )\right)  ,
\end{align}
where $D_\mu = \partial_\mu - i q A_\mu$ is the electromagnetic gauge-covariant derivative.
The obvious diffeomorphism covariant generalization of this expression to curved space is
\begin{align}\label{action}
	S = \int \left( \frac{i}{2} v^\mu \psi^\dagger \overset{\leftrightarrow} D_\mu \psi  - \frac{h^{\mu \nu}}{2m} (D_\mu \psi )^\dagger (D_\nu \psi )\right)
\end{align}
where the integral includes an implicit factor of the volume element $\varepsilon$. For a spinful Schr\"odinger field, the covariant derivative will of course include the spin connection as mentioned above. One immediately retrieves (\ref{flat action}) by going to flat space
\begin{align}
	n_\mu =
	\begin{pmatrix}
		1 & 0 \\
	\end{pmatrix},
	&&
	h^{\mu \nu} =
	\begin{pmatrix}
		0 & 0 \\
		0 & \delta^{ij}
	\end{pmatrix}  ,
\end{align}
and selecting a frame where $v^\mu = \begin{pmatrix} 1 & 0 \end{pmatrix}^T$.

Unfortunately, this action makes explicit reference to $v^\mu$ and so is not manifestly boost invariant. This problem necessarily arises whenever time derivatives are involved since the only diffeomorphism covariant expression that includes a $\partial_0$ is $v^\mu \nabla_\mu$. However, the action can be made to be invariant by introducing a gauge field for mass, 
\begin{align}\label{deriv}
	D_\mu = \partial_\mu - i m \aa_\mu - i q A_\mu
\end{align} 
that transforms in the manner
\begin{align}\label{a transformation}
	\aa_\mu \rightarrow \aa_\mu + k_\mu - \frac{1}{2} n_\mu k^2 .
\end{align}

As presented this may seem ad hoc, however, a direct non-relativistic limit naturally identifies $\aa_\mu$ with the lapse and shift and so corresponds to gravitational forces (see \cite{GPR_geometry}). The transformation (\ref{a transformation}) then naturally follows from an ambiguity in the ADM decomposition of a Lorentzian spacetime. In \cite{Jensen:2014aia}, this gauge field was identified with the electromagnetic field. Effectively, this amounts to collecting $\aa$ and $A$ into the combination $\aa + \frac q m A$, and so the treatment could only describe particles whose mass was tied to their charge. By separating $\aa_\mu$ from $A_\mu$ we have two independent symmetries $U(1)_M$ and $U(1)_Q$ and have the added benefit that we may describe non-relativistic systems whose components have multiple charge to mass ratios. We will exploit this added freedom in section \ref{sec:fluids} to generalize the analysis of \cite{Jensen:2014aia} to fluids that have independent mass and charge currents.

The action (\ref{action}) is boost invariant and a clear generalization of the flat space action. However it is a rather curious combination of one derivative terms, two derivative terms, and frame dependent quantities with precise factors of $m$ to make everything work. It would be nice to know how to systematically generate such combinations at any order in derivatives. To address this problem, we begin by rephrasing (\ref{action}) in a manifestly invariant manner. The result (\ref{covariant schrodinger}) is essentially equivalent to a null reduction and may also be found in equation (3.23) of \cite{Jensen:2014aia}. However, this will teach us how to select the proper Galilean invariant combinations of time and space derivatives at any order we choose.

We shall eventually wish to work with multiple particle species, so let $M$ denote the generator of $U(1)_M$ and Q denote that of $U(1)_Q$, which we will take to be anti-hermitian. For our single field $\psi$ we have for instance $M \psi = i m \psi$ and $M \psi^\dagger = - i m \psi^\dagger$. The derivative
\begin{align}\label{cov div}
	D_\mu = \partial_\mu - \aa_\mu M - A_\mu Q 
\end{align}
is $U(1)_M \times U(1)_Q$ gauge covariant, but not boost covariant and the time and space derivatives $D_0 = v^\mu D_\mu$, $D_a = e^\mu_a D_\mu$ transform as
\begin{align}
	D_0  \rightarrow D_0  + k^a D_a - \frac{1}{2} k^2 M
	&&D_a \rightarrow D_a - k_a M .
\end{align}
These can however be collected into a derivative valued in the dual of the extended representation (\ref{extended})
\begin{align}\label{D extended}
	D_I  =
	\begin{pmatrix}
		D_0  & D_a  & M  
	\end{pmatrix},
	&&\text{so that}
	&&D_I  \rightarrow  (\Lambda^{-1})^J{}_I  D_J .
\end{align}
The obvious two derivative quadratic action is then
\begin{align}\label{covariant schrodinger}
	S = - \frac{1}{2m } \int D_I \psi^\dagger D^I \psi
\end{align}
which upon expanding yields precisely (\ref{action}). The form clearly mirrors that of null compactification, where $g^{IJ}$ plays the role of the higher dimensional metric and (\ref{D extended}) is the derivative operator in that space. The identification $D_{d+1} = i m$ follows by retaining only modes of momentum $m$ in the compactified circle. The advantage of this presentation however is it compactly describes the action (\ref{action}) in a manner intrinsic to the non-relativistic spacetime.

This is a simple rewriting; the true benefit of collecting $D_0 \psi$ and $D_a \psi$ into $D_I \psi$ is that it's now clear how to generalize the precise cancellations in (\ref{action}) to arbitrary boost invariant actions of any order in fields and derivatives. To do so, we extend the definition (\ref{D extended}) to a derivative operator $\mathcal D_I$ on tensors in the extended representation
\begin{align}\label{D def}
	\mathcal D_I u^J = D_I u^J + \omega_I{}^J{}_K u^K ,
	&&\mathcal D_I v_J = D_I v_J -  v_K \omega_I{}^K{}_J ,
\end{align}
and the obvious generalization for more general extended-valued tensors.
By the first term we mean simply apply the operator (\ref{D extended}) to $v_J$.

In the definition (\ref{D def}) we have converted the spacetime index on $\omega_\mu{}^I{}_J$ to an extended index via the projector (\ref{projector})
\begin{align}
	\omega_K{}^I{}_J = \Pi^A{}_K e^\mu_A \omega_\mu{}^I{}_J .
\end{align}
One may then check by hand that $\mathcal D_I$ is locally covariant, which follows in the usual way upon realization that $D_K \Lambda^I{}_J = \Pi^A{}_K e_A^\mu \partial_\mu \Lambda^I{}_J$, giving the required cancellation. For instance
\begin{align}
	\mathcal D_I u^J &= D_I u^J + \omega_I{}^J{}_K u^K \nonumber \\
	&\rightarrow (\Lambda^{-1})^K{}_I D_K ( \Lambda^J{}_L u^L ) + (\Lambda^{-1})^K{}_I \Lambda^J{}_L \omega_K{}^L{}_M u^M + (\Lambda^{-1})^N{}_I \Pi^A{}_N e^\mu_A \Lambda^J{}_L  \partial_\mu (\Lambda^{-1})^L{}_K \Lambda^K{}_M u^M
	\nonumber \\
	&\rightarrow (\Lambda^{-1})^K{}_I \Lambda^J{}_L D_K u^L + (\Lambda^{-1})^K{}_I \Lambda^J{}_L \omega_K{}^L{}_M u^M \nonumber \\
	& \qquad +(\Lambda^{-1})^K{}_I \left( D_K ( \Lambda^J{}_L) u^L  -   \Pi^A{}_K e^\mu_A \partial_\mu  ( \Lambda^J{}_L ) u^L\right)
	\nonumber \\
	&= (\Lambda^{-1})^K{}_I \Lambda^J{}_L \mathcal D_K u^L .
\end{align}

It's now a simple exercise in index contraction to write down Galilean invariant actions of any order $k$ in derivatives. Given a collection of fields $\psi^i$ in some representation of $U(1)_M \times U(1)_Q$ (to account for differing charge/mass ratios), they are of the form
\begin{align}
	S = S [ \psi^ i ,  \mathcal D_I \psi^i , \dots ,  \mathcal D_{I_1} \cdots  \mathcal D_{I_k} \psi^i , \psi^{\dagger}_i ,  \mathcal D_I \psi^{\dagger}_i , \dots ,  \mathcal D_{I_1} \cdots  \mathcal D_{I_k} \psi^{\dagger}_i  ]
\end{align}
where all indices have been contracted using the invariant tensors $g^{IJ}$ and $n_I$ and of course all terms have $U(1)_M \times U(1)_Q$ charge zero.

As an example, consider the term
\begin{align}
	\mathcal D_I \mathcal  D_J \psi^\dagger  \mathcal D^I \mathcal  D^J \psi .
\end{align}
In flat space this gives rise to 
\begin{align}
	D_I D_J \psi^\dagger D^I D^J \psi &= \text{Tr}
	\begin{pmatrix}
		D_0 D_0  \psi^\dagger & D_0 D_c \psi^\dagger & -i m D_0 \psi^\dagger \\
		D_a D_0 \psi^\dagger & D_a D_c \psi^\dagger & -i m D_a  \psi^\dagger \\
		-i m D_0 \psi^\dagger & -im D_c \psi^\dagger & - m^2 \psi^\dagger
	\end{pmatrix}	
	\begin{pmatrix}
		- m^2 \psi & i m D^b \psi & i m D_ 0 \psi \\
		i m D^c \psi & D^c D^b \psi &  D^c D_0  \psi \\
		 i m  D_0 \psi &  D_0 D^b \psi & D_0 D_0  \psi
	\end{pmatrix}\nonumber \\
	&=
	m^2 ( - D_0 D_0 \psi^\dagger \psi + 2 D_0 \psi^\dagger D_0 \psi - \psi^\dagger D_0 D_0 \psi ) \nonumber \\
	&+ i m ( \{ D_0 , D_a \} \psi^\dagger D^a \psi  - D^a \psi^\dagger \{ D_0 , D_a \} \psi  ) + D_a D_b \psi^\dagger D^a D^b \psi 
\end{align}
which one may explicitly check to be invariant under global boosts. The spin connection is not necessary to form globally invariant flat space actions, but is needed to get the proper local boost invariant action, supplying the necessary terms involving derivatives of $v^\mu$ via (\ref{sup}).

\section{Ward Identities}\label{sec:ward}

In preparation for the fluid analysis of section \ref{sec:fluids}, we now turn to the Ward identities associated with diffeomorphism invariance. These merely express local momentum and energy conservation and serve as the equations of motion for thermodynamic variables upon taking the hydrodynamic limit \cite{landau1987fluid}. So far as we can tell, they were first derived in this language in the work of \cite{Janiszewski:2012nb}, which specialized to flat geometries and latter augmented to include a curved metric $h_{ij}$ in \cite{Andreev:2013qsa}.

The full diffeomorphism covariant Ward identities on arbitrary backgrounds were first spelled out in \cite{Geracie:2014nka}, but the treatment was not boost invariant, including a stress tensor and energy current that depended on a choice of $v^\mu$. This was resolved in \cite{Jensen:2014aia}, which first defined a boost invariant, diffeomorphism covariant stress-momentum tensor and stated its Ward identity without reference to $v^\mu$. The work-energy equation however can only be stated in a boost invariant way in the presence of a boost invariant vector field $u^\mu$, for instance in the presence of a fluid \cite{Jensen:2014ama}.

Our approach is essentially equivalent other than issues concerning the connection already discussed, but we take the time to restate these identities in a language that will prove useful to us later on. The only mild innovation is that we collect the stress tensor, energy current, momentum current, and mass current into a single object $\tau^\mu{}_I$ carrying an index valued in the extended representation of the Galilean group. The $v^\mu$ dependent Ward identities then manifest themselves as a single invariant equation (\ref{diff ward}) for work done by external fields on $\tau^\mu{}_I$.

\subsection{The Stress-Energy Tensor}\label{sec:stress-energy}
To do this we begin by comparing (\ref{extended}) and (\ref{a transformation}), noting that the veilbein $e^A_\mu$ and mass gauge field $\aa_\mu$ may be collected into a single object $e^I_\mu$ valued in the extended representation
\begin{align}
	e^I = 
	\begin{pmatrix}
		e^A \\
		\aa
	\end{pmatrix}
\end{align}
which we shall call the ``extended veilbein". As discussed for the derivative operator in section \ref{sec:actions}, one can think of the Newton-Cartan geometry as a null reduction of a Lorentzian geometry in one higher dimension \cite{Jensen:2014aia,GPR_geometry} and the ``extended veilbein'' is simply the reduction of the higher dimensional veilbein. 
Care should be taken with this terminology since there is no sense in which this forms a veilbein on spacetime as the collection is necessarily linearly dependent.

The other background fields are then the Galilean spin connection $\omega^A{}_B$ and the vector potential $A$. We define their associated currents as
\begin{align}
	\delta S = \int \left(-  {\tau^\mu}_I \delta e^I_\mu + {s^{\mu}}_B{}^A \delta{\omega_\mu}^B{}_A + j^\mu \delta A_\mu \right) .
\end{align}
$j^\mu$ is then the charge current and $s^{\mu}{}_B{}^A$ the spin current.  The tensor $\tau^\mu{}_I$ carries an index valued in the extended representation and transforms covariantly. We can then easily retrieve the $v^\mu$ dependent treatments of previous works by parsing out this tensor into it's components.  

The currents as defined previously in the literature \cite{Geracie:2014nka}
\begin{align}
	\delta S = \int \left( \frac{1}{2}  T_\nc^{\mu \nu} \delta h_{\mu \nu} -  \varepsilon_\nc^\mu \delta n_\mu -  p_\mu \delta v^\mu + \rho^\mu \delta \aa_\mu + j^\mu \delta A_\mu \right)
\end{align}
rely essentially on a choice of $v^\mu$, but now are seen to naturally arise in the single object
\begin{align}\label{stress-energy}
	\tau^A{}_I = 
	\begin{pmatrix}
		  \varepsilon_\nc^0 & -   p_{\nc\,b} & - \rho^ 0 \\
		  \varepsilon_\nc^a & -   T_\nc^a{}_{\, b} & - \rho^a_\nc
	\end{pmatrix} .
\end{align}
Here and in what follows the label ``nc'' will be affixed to those currents measured in the ``lab frame'' defined by $v^\mu$.
$ T_\nc^a{}_{\, b}$ is the response to the variation of the purely spatial components of the veilbein and so corresponds to the spatial stress. $ p_{\nc\,b}$ is the momentum current, is purely spatial and as we shall see is equal to the spatial mass current $\rho_{\nc\,b}$ when matter is not charged under boosts, a fact that we will assume in the remainder of this section. We shall refer to $\tau^\mu{}_I$ as the stress-energy tensor for convenience, despite the fact that it contains data beyond energy and stress.

\iffalse
As an example, the stress $\tilde T^{\mu \nu} = e^\mu_a e^\nu_b \tilde T^{ab}$ can be shown to transform as
\begin{align}
	\tilde T^{\mu \nu} \rightarrow \tilde T^{\mu \nu} + {P^\mu}_\lambda  \rho^\lambda k^\nu + k^\mu {P^\nu}_\lambda  \rho^\lambda - k^\mu k^\nu n_\lambda \rho^\lambda
\end{align}
using the transformation matrices (\ref{fundamental}) and (\ref{extended}) applied to $\tau^A{}_I$, matching (5.22) of \cite{Jensen:2014aia} up to an overall sign. Similarly, the energy current $\tilde \varepsilon ^\mu \equiv E^\mu_A \tilde \varepsilon^A$ transforms as
\begin{align}
	\tilde \varepsilon^\mu \rightarrow \tilde \varepsilon^\mu - \tilde T^{\mu \nu} k_\nu + v^\mu  \rho^\nu k_\nu - \frac{1}{2} k^2   \rho^\mu 
\end{align}
matching (5.31) of \cite{Jensen:2014aia}.
As a final point of comparison, we note the covariant stress of \cite{Jensen:2014aia} can be retrieved by proper index shuffling
\begin{align}
	\tau^{AI} \Pi^B{}_I =
	\begin{pmatrix}
		 \rho^0 &  \rho^b \\
		\rho^a & - \tilde T^{ab} 
	\end{pmatrix}.
\end{align}
As noted previously, there is no boost invariant way to isolate the energy current without additional data.

\fi

\subsection{Diffeomorphisms}
In the presence of massive charged fields, the background fields are then an extended veilbein $e^I$, a spin connection $\omega^A{}_B$ and the electromagnetic gauge field $A$.
Their transformations under infinitesimal diffeomorphisms $\xi$ are the standard ones
\begin{align}\label{diffeos}
	&\delta e^I = \iota_\xi d e^I + d \iota_\xi e^I = D \iota_\xi e^I + \iota_\xi T^I - \iota_\xi \omega^I{}_J e^J ,\nonumber \\
	&\delta \omega^A{}_B = \iota_\xi d \omega^A{}_B + d \iota_\xi \omega^A{}_B = \iota_\xi {R^A}_B + D \iota_\xi \omega^A{}_B , \nonumber \\
	&\delta A = \iota_\xi d A + d \iota _\xi A = \iota_\xi F + d \iota_\xi A ,
\end{align}
where $\iota_\xi F$ represents the contraction $\xi^\nu F_{\nu \mu}$ etc.
We have defined an extended torsion tensor
\begin{align}
	T^I = D e^I = d e^I + \omega^I{}_J \wedge e^J = 
	\begin{pmatrix}
		T^A \\
		f
	\end{pmatrix}
\end{align}
$T^I$ includes the full information on the spacetime torsion $T^A$ defined in (\ref{curvTor}) as well as a ``mass" torsion $f = d \aa - \varpi_a \wedge e^a$. We shall see that $f$ couples to the mass current as an external field strength and so should not be thought of as a Newtonian gravitational force, which is encoded within the covariant derivative (see \cite{GPR_geometry} for a more detailed discussion of Newtonian gravity in relation to our geometry).

The transformations (\ref{diffeos}) differ from covariant expressions by a $U(1)_Q$ gauge transformation $- \iota_\xi A$ and an infinitesimal local Galilean transformation $\iota_\xi \omega^A{}_B$. Since our theory is assumed to be gauge and local Galilean invariant we may perform a simultaneous diffeomorphism, $U(1)_Q$ and $Gal(d)$ transformation, under which we have
\begin{align}
	\delta e^I = D \iota_\xi e^I + \iota_\xi T^I
	&&\delta \omega^A{}_B = \iota_\xi R^A{}_B
	&&\delta A = \iota_\xi F.
\end{align}
Variation of the action then yields the Ward identity
\begin{align}\label{diff ward}
	- e^I_\mu ( D_\nu - {T^\lambda}_{\lambda \nu}) {\tau^\nu}_I + {T^I}_{\mu \nu} {\tau^\nu}_I = F_{\mu \nu} j^\nu + {R^B}_{A\mu \nu} {s^{\nu}}_B{}^A  .
\end{align}
As ever, $D_\mu$ represents the locally Galilean covariant derivative $D_\mu \tau^\nu{}_I = \nabla_\mu \tau^\nu{}_I - \tau^\nu{}_J \omega_\mu{}^J{}_I$.

One might worry about the appearance of $\aa_\mu$ in the first term. However, if mass is conserved (as we shall always assume), the explicit $\aa_\mu$ term drops out, as it multiplies the Ward identity
\begin{align}
	( \nabla_\mu - T^\nu{}_{\nu \mu} ) \rho^\mu = 0,
\end{align}
which follows in the usual way upon a $U(1)_M$ gauge variation $\delta a = d \Lambda_M$. Of course, charge conservation follows as well from $\delta A = d \Lambda_Q$
\begin{align}
	( \nabla_\mu - T^\nu{}_{\nu \mu} ) j^\mu = 0.
\end{align}
Despite the deceptive appearance of the mass gauge field, we shall find (\ref{diff ward}) very useful owing to it's manifest boost invariance.

Equation (\ref{diff ward}) contains both the work-energy equation and momentum conservation equation of previous works. To get a sense of this equation, the reader may find (\ref{parsed ward}) helpful, where these components are isolated  and rendered in a more familiar form for fluid systems. Finally, we note that since we have defined the stress and spin currents in a veilbein formalism, our stress is a response to a shear at fixed $\omega_\mu{}^A{}_B$, which necessarily introduces torsion $T^A = D e^A$. The physical stress tensor rather measure response to a shear at fixed spatial torsion. The proper redefinition of currents will introduce additional terms into the Ward identity involving double derivatives of the spin current and so will not enter into our calculations with the power counting scheme we will adopt. For this reason we shall also drop the final term in the Ward identity.

\subsection{Local Boosts}\label{sec:boost ward}
It is well known that Galilean symmetry fixes the momentum to the flow of matter. 
The relevant Ward identity to demonstrate this is that of infinitesimal local boosts, under which we have
\begin{align}
	\delta \aa =  k_a e^a ,
	&&\delta e^a = - k^a e^0 , 
	&& \delta \varpi^a = D k^a = d k^a + \omega^a{}_b k^b ,
\end{align}
while the spin connection $\omega^a{}_b$ is neutral. Variation of the action then gives
\begin{align}
	0 = \int \left(   \rho_\nc^a k_a - p_\nc^a k_a + s^{\mu}{}_a{}^0 D k^a\right) ,
\end{align}
from which we have 
\begin{align}
	 p_\nc^a =  \rho_\nc^a - ( D_\mu - {T^\nu}_{\nu \mu} )s^{\mu 0 a } .
\end{align}

Interestingly we see that the oft-cited non-relativistic identity between the mass and momentum currents may be violated in the presence of matter that couples to the boost part of the spin connection $\varpi^a$
\begin{align}
	D_\mu = \partial_\mu  - \aa_\mu M  - A_\mu Q- \varpi_\mu{}^a K_a + \frac{1}{2} \omega_\mu{}^{ab} J_{ab} .
\end{align}
Here $J_{ab}$ are the generators of rotations and $K_a$ of boosts, which, together with translations $P_a$, time translations $H$ and mass $M$, close to form the Bargmann algebra \cite{Bargmann:1954gh}
\begin{align}\label{BargmannAlgebra}
	&[ J^{ab} , J^{cd} ] = i \left( \delta^{ac} J^{bd} - \delta^{ad} J^{bc} - \delta^{bc} J^{ad} + \delta^{bd} J^{ac} \right), \nonumber \\
	&[J^{ab} , P^c ] = i \left( \delta^{ac} P^b - \delta^{bc} P^a \right),
	&&[J^{ab} , K^c ] = i \left( \delta^{ac} K^b - \delta^{bc} K^a \right), \nonumber \\
	&[P^a , K^b] = - i\delta^{ab} M,
	&& [ H , K^a ] = - i P^a .
\end{align}

The identity $ p_\nc^a =  \rho_\nc^a$ is then generically violated whenever there is matter that transforms under the sub-algebra spanned by $\{ J_{ab} , K_a \}$ in which $K_a$ is represented non-trivially
\begin{align}
	\psi \rightarrow e^{ i k^a K_a} e^{- \frac{i}{2} \theta^{ab} J_{ab} }\psi .
\end{align}
Such representations were first considered in the work of Levy-Leblond \cite{LevyLeblond:1967zz} and enumerated up to and including spin 1 in \cite{0305-4470-39-29-026,Niederle:2007xp}. However, since we do not know of any condensed matter systems that realize these representations we shall assume $K_a =0$ in what follows so that $s^{\mu a 0} = 0$ and the identity between momentum and mass currents is retained
\begin{align}
	 p_\nc^a = \rho_\nc^a .
\end{align} 

\section{Non-Relativistic Fluids}\label{sec:fluids}

In this section we turn to non-relativistic fluid dynamics to illustrate the advantages of our approach. The program of fluid dynamics is to describe the fluctuations of thermodynamic variables in slightly out of equilibrium media. These variables include the temperature and velocity $u^\mu$ as well as a chemical potential for each conserved charge. For us, there are two such quantities, the electric charge and mass, whose associated chemical potentials we will denote $\mu_Q$ and $\mu_M$ ($\mu_M$ being the same boost invariant $\mu_M$ defined in \cite{Jensen:2014ama}). In all we have $d+3$ degrees of freedom
\begin{align}\label{therm var}
	T , 
	&&\mu_Q , 
	&&\mu_M , 
	&&u^\mu,
\end{align}
where the velocity has been normalized so that $n_\mu u^\mu = 1$.
The equilibrium properties of the system are then completely characterized by an equation of state such as $p(T, \mu_Q , \mu_M )$, which expresses the pressure as a function of the state variables. The entropy, charge, mass, and energy densities are then defined by
\begin{align}\label{thermo}
	dp = s dT + q d \mu_Q + \rho d \mu_M,
	&&\epsilon + p = T s + q \mu_Q + \rho \mu_M .
\end{align}

Just out of equilibrium, the thermodynamic variables are allowed to vary slowly in space and time, where slow is compared to the mean free path and mean free time so that a local equilibrium is always a good approximation. The Ward identities 
\begin{align}\label{fluid eom}
	&(\nabla_\mu - {T^\nu}_{\nu \mu} ) j^\mu = 0 , \qquad \qquad
	(\nabla_\mu - {T^\nu}_{\nu \mu} ) \rho^\mu = 0 , \nonumber \\
	&- e^I_\mu ( D_\nu - {T^\lambda}_{\lambda \nu}) {\tau^\nu}_I + {T^I}_{\mu \nu} {\tau^\nu}_I = F_{\mu \nu} j^\nu   ,
\end{align}
are then sufficient to serve as equations of motion since they are the same in number as the thermodynamic variables. We need only constitutive relations for the currents in terms of these degrees of freedom and their derivatives.

Since we are perturbing away from equilibrium, these constitutive relations naturally organize themselves in a gradient expansion where higher derivatives of (\ref{therm var}) take on diminishing importance. To complete our power counting scheme one needs also specify the backgrounds to be perturbed around. In this paper we shall assume a trivial background in equilibrium so that 
\begin{align}
	F_{\mu \nu}, 
	&&\nabla_\mu,
	&&T^I{}_{\mu \nu}
\end{align}
are all first order in derivatives. This is not a unique selection and corresponds to a choice of regime in which we expect our results to be applicable. One may for example consider backgrounds of large constant magnetic field in which $B$ appears at zeroth order and the analysis will be significantly altered.

The constitutive relations so obtained are not in general sensible and could lead to an on-shell decrease in entropy, the canonical example being that of a negative shear viscosity \cite{landau1987fluid}. As such we must also impose the second law of thermodynamics as an additional constraint on fluid flows, diminishing the freedom present in the gradient expansion and returning a reduced set of transport coefficients. This procedure has been carried out in many systems including  $2+1$ dimensional relativistic normal fluids \cite{Jensen:2011xb} and $3+1$ dimensional relativistic superfluids \cite{Bhattacharya:2011tra}.

In this paper we will work only to first order in derivatives to illustrate our method in the non-relativistic setting. It's well known that even at first order, the dynamics is very rich in the parity odd sector when $d=2$ and so we shall eventually restrict to two dimensional fluids.

Although our calculation is very much along the lines of \cite{Jensen:2014ama}, there are two key differences. The first is the existence of an independent mass current, a necessary element for treating systems with multiple components. One might for example consider mixtures of He-3 and He-4 in their normal phases.  It shouldn't be surprising that such systems admit  a richer transport sector. In this work we find that a multicomponent fluid admits one additional sign semi-definite transport coefficient, one additional unconstrained parity even coefficient and three additional parity odd coefficients compared to the single component fluid. A succinct overview of these results may be found in section \ref{sec:single component}.
Our second point of divergence is that stressed in section \ref{sec:intro} and we will be using an connection without the kinematical restrictions imposed in previous works.

\subsection{Covariant Currents}
As we've already noted, the currents come naturally assembled into a single stress-energy tensor $\tau^\mu{}_I$. However, in the presence of a boost invariant fluid velocity $u^\mu$ we may go further and define Galilean frame invariant notions of energy and stress separately. Heuristically, this corresponds to defining them to be as measured in a frame co-moving with $u^\mu$. In this section, we demonstrate the details of how to do this. Our approach in this regard is essentially equivalent to that found in \cite{Jensen:2014ama}, though we restate it here in our language.
These currents at hand, we parse the Ward identity (\ref{diff ward}) into something a bit more familiar: the work-energy and Navier-Stokes equations. 

To begin, consider the fluid velocity as measured with respect to some lab frame 
\begin{align}
	u^A \equiv e^A_\mu u^\mu =
	\begin{pmatrix}
		1 \\
		u^a
	\end{pmatrix} .
\end{align}
The existence of a preferred $u^A$ allows us to define $P^A{}_B = \delta^A{}_B - u^A n_B $ and a $h_{AB}$ such that $h^{AC} h_{CB} = P^A{}_B$ unambiguously\footnote{In this section only, these tensors correspond to the fluid frame, not the frame defined by $v^\mu$.}. We shall also make use of
\begin{align}
	P^A{}_I = P^A{}_B \Pi^B{}_I =
	\begin{pmatrix}
		0 & 0 & 0\\
		- u^a & \delta^a{}_b & 0 
	\end{pmatrix}
\end{align}
which projects extended indices to transverse vector indices.

Any vector index be decomposed uniquely into a part parallel to $u^A$ and perpendicular to $n_A$, while an dual vector index may be decomposed into a part parallel to $n_A$ and perpendicular to $u^A$. For instance, for a vector $v^A$ and covector $w_A$ we may write
\begin{align}
	v^A = v u^A + v'^A,
	&&w_A = w n_A + w'_A
\end{align}
where $v'^A n_A = u^A w'_A = 0$.

We should like to perform a similar decomposition for extended indices $I$ for which we have the preferred vector
\begin{align}
	n^I = g^{IJ} n_J = 
	\begin{pmatrix}
		0 \\
		0 \\
		1
	\end{pmatrix}.
\end{align}
 However, $n^I$ is a null vector of the Lorenztian metric $g_{IJ}$ and is thus perpendicular to itself, making the above procedure impossible. In Lorentzian geometry one usually continues by introducing a second null vector $l^I$ whose inner product with $n^I$ is 1, though such a $l^I$ is not unique. One can then decompose any index into parts parallel to $n^I$, parallel to $l^I$, and perpendicular to both. Luckily in the presence of a background fluid, there is a preferred way to select such a vector. We will define $u^I$ in the extended representation so that is both null $u_I u^I = 0$ and projects to the fluid velocity in the sense $u^A = \Pi^A{}_I u^I$. In components, the $u^I$ so defined is of the form
\begin{align}
	u^I = 
	\begin{pmatrix}
		1 \\
		u^a \\
		- \frac{1}{2} u^2
	\end{pmatrix}
\end{align} 
and automatically satisfies $n_I u^I = 1$. 

Using this extended velocity vector, we now decompose the lower index of $\tau^A{}_I$ into parts parallel to $n_I$, parallel to $u_I$ and perpendicular to both
\begin{align}
	\tau^A{}_I &= \varepsilon^A n_I - \rho^A u_I + t^A{}_I ,
\end{align}
where by definition $t^A{}_I n^I = t^A{}_I u^I = 0$. When an $I,J, \dots$ index is transverse in this sense, it may always be written as the pullback of a unique tensor with a transverse $A,B, \dots$ indices. In our case we have $t^A{}_I = t^A{}_B \Pi^B{}_I$ where $t^A{}_B u^B = 0$. We now continue, decomposing the upper index of $t^A{}_B$ into a part parallel to $u^A$ and a part perpendicular to $n_A$
\begin{align}
	t^A{}_B = u^A p_B + T^A{}_B .
\end{align}

Altogether we have
\begin{align}\label{fluid currents}
	\tau^A{}_I &= \varepsilon^A n_I - \rho^A u_I - ( u^A p_B + T^A{}_B  ) \Pi^B{}_I .
\end{align}
We can invert this definition to find
\begin{align}
	\varepsilon^A = \tau^A{}_I u^I ,
	&& \rho^A = - \tau^A{}_I n^I ,
	&&p^A =- n_B \tau^B{}_I P^{AI},
	&&T^{AB} = - {P^A}_C \tau^C{}_I   P^{BI}.
\end{align}
These are the boost invariant currents we shall use in the fluid analysis and can be so defined whenever there exists a preferred velocity vector to draw on. They correspond to the energy, mass current, momentum and stress as measured by an observer comoving with the fluid. The relation to the component decomposition (\ref{stress-energy}) is
\begin{align}\label{nc currents}
	&~ \qquad \varepsilon^A = 
	\begin{pmatrix}
		 \varepsilon_\nc^0 -  \rho_\nc^b u_b + \frac{1}{2} \rho^0 u^2 \\
		 \varepsilon_\nc^a -  T_\nc^{ab} u_b + \frac{1}{2}  \rho^a_\nc u^2
	\end{pmatrix},
	&&\qquad  \qquad  \rho^A = 
	\begin{pmatrix}
		 \rho^0 \\
		  \rho_\nc^a
	\end{pmatrix}, \nonumber \\
	&T^{AB} =
	\begin{pmatrix}
		0 & & 0 \\
		0 & &  T_\nc^{ab} + u^a   \rho_\nc^b +   \rho_\nc^a u^b -  \rho^0 u^a u^b 
	\end{pmatrix} ,
	&&p^A = 
	\begin{pmatrix}
		0 \\
		 \rho_\nc^a -  \rho^0 u^a
	\end{pmatrix}
	= P^A{}_B \rho^B.
\end{align}
where we have used the Ward identity $ p_\nc^a =  \rho_\nc^a$.  We note that $\varepsilon^A$ is simply the Milne covariant energy current defined in \cite{Jensen:2014ama}.

We now turn to restating the Ward identity (\ref{diff ward}) in terms of these currents. To simplify matters, we shall now take $v^\mu = u^\mu$ so that $u^A = \begin{pmatrix} 1 & 0 \end{pmatrix}^T$ and define the transverse projector and inverse metric accordingly.

A few words on torsion are due before we proceed.  Throughout we shall restrict to backgrounds in which there is no spatial torsion. Spatial torsion is necessary in the study of media with dislocation defects, where the presence of torsion simply indicates a nonzero Burgers vector \cite{Hughes:2012vg,Geracie:2014mta} (for a classic discussion of dislocations in elastic media see \cite{landau1986theory}) and would be an interesting element to include in future work. For our purposes however, the fluid analysis is greatly simplified by discarding it.

The minimal temporal torsion (\ref{temporal torsion}) may also be set to zero by assuming the clock-form is closed. Although this is the case in all physically relevant situations, temporal torsion is useful for studying energy transport \cite{Geracie:2014zha} as it is equivalent to coupling the system to a Luttinger potential \cite{Luttinger:1964zz}.
As noted previously, our formalism also allows for a non-zero ``mass" torsion $f= d\aa - \varpi^a \wedge e_a$ which will be zero on physical backgrounds. Although including mass torsion in our analysis will not prove as fruitful as temporal torsion, there is no essential difficulty in doing so and so we keep it around.

The boost covariant manner in which we set spatial torsion to zero but keep these effects is a decomposition of the extended torsion tensor $T^I$ along the lines of the previous section, keeping only components parallel to $n^I$ and $u^I$
\begin{align}
	T^I = u^I G + n^I \ff
\end{align}
where $G= dn$ and $\ff =  f + u_a T^a - \frac{1}{2} u^2 G$. This decomposition would in general obtain a third term that is both $n$ and $u$ orthogonal corresponding to spatial torsion but which we have set to zero here.
The spacetime torsion tensor is then ${T^\lambda}_{\mu \nu} = u^\lambda G_{\mu \nu}$, whose pullback to a slice is zero if we assume $n \wedge dn = 0$.
The Ward identity for diffeomorphisms then becomes
\begin{align}
	- e^I_\mu ( D_\nu - G_\nu) {\tau^\nu}_I = \ff_{\mu \nu} \rho^\nu + F_{\mu \nu} j^\nu - G_{\mu \nu} \varepsilon^\nu .
\end{align}
We see that $G$ serves as a field strength coupled to energy and $\ff$ a field strength coupled to mass. We have also defined $G_\mu = - G_{\mu \nu} u^\nu = T^\nu{}_{\nu \mu}$.

Now supplement the decomposition (\ref{fluid currents}) of the stress-energy with one of the extended veilbein
\begin{align}\label{veilbein decomposition}
	e^I_\mu = n^I c_\mu + u^I b_\mu + q^I n_\mu + {q^I}_\mu
\end{align}
where by definition
\begin{align}
	&u_I q^I = n_I q^I = 0,
	&&u_I {q^I}_\mu = n_I {q^I}_\mu = 0,
	&&{q^I}_\mu u^\mu = 0 .
\end{align}
A straightforward computation then shows that
\begin{align}
	c_\mu = \aa_\mu 
	&&b_\mu = n_\mu,
	&&q^I = 0,
	&&q^I{}_\mu = 
	\begin{pmatrix}
		e^A_\mu - u^A n_\mu \\
		0 
	\end{pmatrix} .
\end{align}

The $\aa_\mu$ term drops out of the equation of motion as it multiplies the continuity equation for mass.
The rest then reads
\begin{align}\label{intermediate eqn}
	&- ( u^I n_\mu + q^I{}_\mu ) ( D_\nu - G_\nu ) \tau^\nu{}_I  = \ff_{\mu \nu} \rho^\nu + F_{\mu \nu}  j^\nu - G_{\mu \nu } \varepsilon^\nu .
\end{align}
Contracting with $u^\mu$ we have
\begin{align}
	&u^I ( D_\mu - G_\mu ) \tau^\mu{}_I =  e_\mu \rho^\mu + E_\mu j^\mu + G_\mu \varepsilon^\mu \nonumber \\
	\implies \qquad & (\nabla_\mu - G_\mu ) \varepsilon^\mu =  e_\mu \rho^\mu + E_\mu j^\mu + G_\mu \varepsilon^\mu +  \tau^\nu{}_I D_\nu u^I .
\end{align}
where
\begin{align}
	e_\mu = \ff_{\mu \nu} u^\nu , &&
	E_\mu = F_{\mu \nu} u^\nu .
\end{align}
are the field strengths observed by co-moving observers.
The final term is rather mysterious looking but can be easily evaluated
\begin{align}\label{manipulations}
	\tau^\nu{}_I D_\nu u^I &= ( \varepsilon^\nu n_I - \rho^\nu u_I - ( u^\nu p_A + {T^\nu}_A ) \Pi^A{}_I ) D_\nu u^I \nonumber \\
	&= -( u^\nu p_A + {T^\nu}_A ) \Pi^A{}_I D_\nu u^I \nonumber \\
	&= -( u^\nu p_A + {T^\nu}_A ) D_\nu u^A \nonumber \\
	&= -( u^\nu \rho_\lambda + {T^\nu}_\lambda ) \nabla_\nu u^\lambda \nonumber \\
	&= -\rho_\nu \alpha^\nu - \frac{1}{2} \sigma^{\nu \lambda} T_{\nu \lambda} - \frac{1}{d} \theta g^{\nu \lambda} T_{\nu \lambda}.
\end{align}
In the second line we have used that $D_\nu n_I = 0$, $n_I u^I = 1$ and $u_I D_\nu u^I = 0$ and in the third line that $D_\nu \Pi^A{}_I = 0$.
The acceleration, shear and expansion of the fluid appearing in the above formula are defined as
\begin{align}
	&\alpha^\mu = u^\nu \nabla_\nu u^\mu ,
	&&\sigma^{\mu \nu} = \nabla^\mu u^\nu + \nabla^\nu u^\mu - \frac{2}{d} g^{\mu \nu} \theta, 
	&&\theta = \nabla_\mu u^\mu .
\end{align}

Altogether
\begin{align}
	(\nabla_\mu - G_\mu ) \varepsilon^\mu =  (e_\mu - \alpha_\mu ) \rho^\mu + E_\mu j^\mu + G_\mu \varepsilon^\mu - \frac{1}{2} \sigma^{\mu \nu} T_{\mu \nu} - \frac{1}{d} \theta g^{\mu \nu} T_{\mu \nu}
\end{align}
which is the work-energy equation, including work done by the external fields as well as dissipated by fluid shears. This matches exactly the covariant work-energy equation as it appears in \cite{Geracie:2014nka} and \cite{Jensen:2014aia} except for the additional mass current that couples to $\ff_{\mu \nu}$.

To obtain the Navier-Stokes equation we raise the $\mu$ index on (\ref{intermediate eqn})
\begin{align}
	- (\nabla_\nu - G_\nu ) ( \tau^\nu{}_I q^{I\mu}) = \ff^\mu{}_{\nu} \rho^\nu + F^\mu{}_{\nu}  j^\nu - G^\mu{}_{\nu } \varepsilon^\nu -  \tau^\nu{}_I D_\nu q^{I\mu} .
\end{align}
The stress tensor term on the LHS is evaluated as
\begin{align}
	- \tau^\nu{}_I q^I{}_\mu &= ( u^\mu p_A + T^\mu{}_A  ) \Pi^A{}_I q^I{}_\mu  = u^\nu \rho_\mu + {T^\nu}_\mu .
\end{align}

Similar manipulations to (\ref{manipulations}) gives
\begin{align}
	- \tau^\nu{}_I D_\nu q^{I\mu} = \rho^\nu u_I D_\nu q^{I\mu} + ( u^\nu \rho_\lambda + T^\nu{}_\lambda ) \nabla_\nu ( e^\lambda_A \Pi^A{}_I q^{I\mu} ) .
\end{align}
Now $e^\lambda_A \Pi^A{}_I q^{I\mu} = h^{\lambda \mu}$ so that the final term drops out.
To evaluate the first term we have
\begin{align}
	u_I D_\nu q^{I\mu} &= u_I ( \nabla_\nu q^{I \mu } + \omega_\nu{}^I{}_J q^{J\mu} ) 
	= - q^{I \mu} \nabla_\nu u_I + u_I \omega_\nu{}^I{}_J q^{J\mu } \nonumber\\
	&= 0 + 
	\begin{pmatrix}
		0 & 0 & 1
	\end{pmatrix}
	\begin{pmatrix}
		0 & 0 & 0 \\
		\varpi_\nu{}^a & \omega_\nu{}^a{}_b & 0 \\
		0 & -\varpi_{\nu b} & 0 
	\end{pmatrix}
	\begin{pmatrix}
		0 \\
		e^{b\mu} \\
		0
	\end{pmatrix} 
	= - e^\mu_a\varpi_\nu^a = - \nabla_\nu u^\mu .
\end{align}

Rearranging the vector equation of motion and using the mass conservation Ward identity we have
\begin{align}\label{NS}
	 &( \nabla_\nu - G_\nu ) (  u^\nu P^\mu{}_\lambda \rho^\lambda + T^{\mu \nu} ) = {\ff^\mu}_\nu \rho^\nu + {F^\mu}_\nu j^\nu - {G^\mu}_\nu \varepsilon^\nu - \rho^\nu \nabla_\nu u^\mu \nonumber \\
	  \implies \qquad &( \nabla_\nu - G_\nu ) ( \rho^\mu u^\nu + u^\mu \rho^\nu - \rho u^\mu u^\nu + T^{\mu \nu} ) = {\ff^\mu}_\nu \rho^\nu + {F^\mu}_\nu j^\nu - {G^\mu}_\nu \varepsilon^\nu 
\end{align}
where $\rho = n_\lambda \rho^\lambda$.

In summary, we may restate the Ward identities in terms of the covariant currents as a work-energy equation and Navier-Stokes equation, plus conservation laws
\begin{align}\label{parsed ward}
	& (\nabla_\mu - G_\mu ) \varepsilon^\mu = (e_\mu - \alpha_\mu ) \rho^\mu + E_\mu j^\mu + G_\mu \varepsilon^\mu - \frac{1}{2} \sigma^{\mu \nu} T_{\mu \nu} - \frac{1}{d} \theta g^{\mu \nu} T_{\mu \nu} \nonumber \\
	 &( \nabla_\nu - G_\nu ) ( \rho^\mu u^\nu + u^\mu \rho^\nu - \rho u^\mu u^\nu + T^{\mu \nu} ) = {\ff^\mu}_\nu \rho^\nu + {F^\mu}_\nu j^\nu - {G^\mu}_\nu \varepsilon^\nu  , \nonumber \\
	 & ( \nabla_\mu - G_\mu ) \rho^\mu = 0 , \qquad \qquad \qquad \qquad
	 ( \nabla_\mu - G_\mu  ) j^\mu  = 0 .
\end{align}

\subsection{Perfect Fluids}
The Ward identities (\ref{parsed ward}) serve as dynamical equations once constitutive relations have been supplied, specifying the currents $j^\mu$, $\rho^\mu$, $\varepsilon^{\mu}$ and $T^{\mu \nu}$ in terms of the thermodynamic degrees of freedom
\begin{align}
	T,
	&&\mu_Q,
	&&\mu_M,
	&&u^\mu.
\end{align}
At zeroth order in derivatives, the most general tensors we can construct using these variables and the Newton-Cartan structure are
\begin{align}
	j^\mu = q u^\mu,
	&&\rho^\mu = \rho u^\mu,
	&&\varepsilon^\mu = \epsilon u^\mu,
	&&T^{\mu \nu} = p h^{\mu \nu} .
\end{align}
Here $q$, $\rho$, $\epsilon$ and $p$ are functions of $(T, \mu_Q , \mu_M)$. They are identified with the thermodynamic charge density, mass density, energy density and pressure and so satisfy the relations (\ref{thermo}).

Feeding these into the Ward identities (\ref{parsed ward}) we obtain the perfect fluid equations of motion
\begin{align}\label{Navier}
	&\dot q + q \theta = 0 ,
	\qquad \qquad
	 \dot \rho + \rho \theta = 0 ,
	\qquad \qquad
	\dot \epsilon + ( \epsilon + p ) \theta = 0 , \nonumber \\
	& \rho \alpha^\mu = \rho e^\mu + q E^\mu + ( \epsilon + p ) G^\mu - \nabla^\mu p .
\end{align}
In these equations and those that follows dotted objects indicate the material derivative, $\dot f = u^\mu \nabla_\mu f$. 
The final equation is simply Newton's second law and an obvious covariant generalization of Euler's equation \cite{landau1987fluid}. It expresses the fact that fluid particles will tend flow along geodesics, deviating only due to the exertion forces from external background fields and internal pressure.

%We note that even the perfect fluid equations of motion differ in our approach from those found in \cite{Jensen:2014ama} and so we would not expect to obtain similar results.

\iffalse
Our equations of motion differ significantly from those found in \cite{Jensen:2014ama} due to the extra kinematic constraint $u^\nu \nabla_\nu u^\mu = E^\mu$ found therein. Specializing to a single component fluid so that $\rho = q$, adopting this connection would restrict the Navier-Stokes equation to be pure constraint
\begin{align}
	0 = \rho e^\mu + ( \epsilon + p ) G^\mu - \nabla^\mu p , \qquad
	\implies \qquad e^\mu - T \nabla^\mu \left( \frac {\mu_M} T \right) = \frac{\epsilon + p }{T \rho} \left( \nabla^\mu T - T g^\mu \right)
\end{align}
which is exactly (4.35) of \cite{Jensen:2014ama} when mass and charge are identified.

\fi

\subsection{Fluid Frames}
To go beyond perfect fluids one needs to expand the currents to first order in derivatives. The constitutive relations are then the perfect fluid ones plus $\mathcal O (\partial^1 )$ corrections
\begin{align}\label{corrections}
	&j^\mu = ( q + \mathcal Q ) u^\mu + \nu^\mu, 
	&& \rho^\mu = ( \rho + \varrho ) u^\mu + \mu^\mu , \nonumber \\
	& \varepsilon^\mu = ( \epsilon + \mathcal E ) u^\mu + \xi^\mu , 
	&&T^{\mu \nu} = ( p + \mathcal P ) h^{\mu \nu} + \pi^{\mu \nu}.
\end{align}
In the above the vector corrections are defined to be transverse
\begin{align}
	n_\mu \nu^\mu = n_\mu \mu^\mu = n_\mu \xi^\mu = 0 
\end{align}
while the tensor correction is traceless
\begin{align}
	h^{\mu \nu} \pi_{\mu \nu}
\end{align}
($\pi^{\mu \nu}$ is of course already transverse since $T^{\mu \nu}$ is). This is convenient as it separates the first order corrections into irreducible representations of $SO(d)$.

The decomposition (\ref{corrections}) is subject to a well known ambiguity stemming from the need to define $T$, $\mu_Q$, $\mu_M$ and $u^\mu$ out of equilibrium. Any such definition is admissible so long as it reduces to the equilibrium values at order zero and so is subject to a $d+3$ parameter $\mathcal O( \partial^1 )$ field redefinition
\begin{align}
	T \rightarrow T + \delta T ,
	&& \mu_Q \rightarrow \mu_Q + \delta \mu_Q,
	&& \mu_M \rightarrow \mu_M + \delta \mu_M,
	&&u^\mu \rightarrow u^\mu + \delta u^\mu
\end{align}
called a fluid frame transformation (not to be confused with a Galilean frame transformation). To deal with this ambiguity we may either fix the frame by imposing extra conditions, or work in a manifestly frame invariant manner. 

Frame transformations are worked out in \cite{Jensen:2014ama} (see \cite{Bhattacharya:2011eea
} for a relativistic treatment) and in this section and the next, we refer the reader to this treatment for the details. For our purposes we only note that (besides those related to the entropy) the complete set of first order frame invariants is
\begin{align}\label{frame invariants}
	&\mathcal S  = \mathcal P - \partial_\epsilon p \mathcal E - \partial_q p \mathcal Q - \partial_ \rho p \varrho,
	\qquad \qquad
	\mathcal T^{\mu \nu} = \pi^{\mu \nu} ,\nonumber \\
	&\mathcal J^\mu = \nu^\mu - \frac{q}{\rho} \mu^\mu, \qquad \qquad
	\mathcal E^\mu = \xi^\mu - \frac{\epsilon + p}{\rho} \mu^\mu .
\end{align}
 Although we shall usually take $p$ to be a function of temperature and the chemical potentials, here we have taken $p = p(\epsilon , q , \rho )$ and the partial derivatives $\partial_\epsilon$, $\partial_q$ and $\partial_\rho$ are defined accordingly.
Note we have an additional vector frame invariant compared to either the relativistic case or non-relativistic single-component fluids  due to the presence of the conserved current $\rho^\mu$.

\subsection{The Entropy Current}
It is convenient to separate out the first order entropy current into a ``canonical part'' and corrections.
Here the canonical part is defined to be that combination of currents chosen to match the equilibrium identity (\ref{thermo})
\begin{align}
	 & T s^\mu_\text{can} = p u^\mu + \varepsilon^\mu - \mu_Q j^\mu - \mu_M \rho^\mu \nonumber \\
	\text{i.e.} \qquad &s^\mu_\text{can} = s u^\mu - \frac{\mu_Q}{T} ( \mathcal Q u^\mu + \nu^\mu ) - \frac{\mu_M}{T} ( \varrho u^\mu + \mu^\mu ) + \frac{1}{T} ( \mathcal E u^\mu + \xi^\mu ) .
\end{align}
Out of equilibrium the entropy flow will in general deviate from the canonical part
\begin{align}
	s^\mu = s^\mu_\text{can} + \zeta^\mu .
\end{align}
This separation is helpful since $s^\mu_\text{can}$ is a frame invariant, and so $\zeta^\mu$ is as well. It's divergence is a quadratic form in first order data 
\begin{align}
	( \nabla_\mu - G_\mu ) s^\mu_\text{can} &=  - \frac{1}{T} \mathcal S \theta  - \frac{1}{2 T} \sigma_{\mu \nu} \mathcal T^{\mu \nu}
	+ \frac{1}{T} \mathcal J^\mu \left( E_\mu - T \nabla_\mu \left( \frac{\mu_Q}{T} \right)  \right) - \frac{1}{T^2} \mathcal E^\mu  ( \nabla_\mu T - T G_\mu )  .
\end{align}

\subsection{Constitutive Relations}
All the necessary tools are now available to carry out the analysis outlined at the beginning of this section: first write out the most general constitutive relations for the first order frame invariants and then impose the second law of thermodynamics. We shall take $d=2$ throughout. The formulae will prove somewhat simpler if we instead take our independent variables to be
\begin{align}
	&T,
	&&\nu_Q = \mu_Q /T ,
	&&\nu_M = \mu_M /T .
\end{align}
In terms of $\nu_Q$ and $\nu_M$, the thermodynamic identities read
\begin{align}\label{thermo2}
	d p = \frac{\epsilon + p}{T} dT + T q d \nu_Q + T \rho d \nu_M,
	&&\frac{\epsilon + p}{T} =  s + \nu_Q q + \nu_M \rho .
\end{align}

Now consider the available first order data
\begin{align}
	\nabla_\mu T,
	&&\nabla_\mu \nu_Q,
	&& \nabla_\mu \nu_M,
	&&\nabla_\mu u^\nu,
	&&\ff_{\mu \nu},
	&&F_{\mu \nu},
	&&G_{\mu \nu} .
\end{align}
Separating into irreducible representations of $SO(2)$ we have
\begin{center}
\begin{tabular}{ l | c c c}
		& Data 	  \\
	Scalar & $\theta \qquad b \qquad B \qquad \omega  $  \\
	&$( \dot T) \qquad ( \dot\nu_Q)  \qquad ( \dot\nu_M)$   \\
	Vector & $ \nabla^\mu T \qquad \nabla^\mu \nu_Q \qquad \nabla^\mu \nu_M \qquad (\alpha^\mu)$\\
		& $ e^\mu \qquad  E^\mu  \qquad G^\mu$ \\
	Symmetric Traceless Tensor & $\sigma^{\mu \nu}$ \\
\end{tabular}
\end{center}
where 
\begin{align}
	&E^\mu = {F^\mu}_\nu u^\nu 
	&&B = \frac{1}{2} \varepsilon^{\mu \nu} F_{\mu \nu} \nonumber \\
	&e^\mu = {\ff^\mu}_\nu u^\nu 
	&&b = \frac{1}{2} \varepsilon^{\mu \nu } \ff_{\mu \nu } \nonumber \\
	&G^\mu = - {G^\mu}_\nu u^\nu 
	&&\theta = \nabla_\mu u^\mu \nonumber \\
	&\sigma^{\mu \nu} = \nabla^\mu u^\nu + \nabla^\nu u^\mu - g^{\mu \nu} \theta
	&&\omega = \varepsilon_{\mu \nu \lambda} u^\mu \nabla^\nu u^\lambda \nonumber \\
	&\alpha^\mu = u^\nu \nabla_\nu u^\mu .
\end{align}
Recall that by $\varepsilon^{\mu \nu}$ we mean the ``spatial volume element'' $\varepsilon^{\mu \nu \lambda} n_\lambda$, which is boost invariant.
Not all this data is independent on-shell. We may thus use the Navier-Stokes equation to eliminate one vector degree of freedom and one scalar each for mass conservation, charge conservation and the work-energy equation. The eliminated data is indicated by parentheses in the above table.

The above amounts to a decomposition of $\nabla_\mu u^\nu$ of the form
\begin{align}\label{decomposition}
	\nabla_\mu u^\nu = n_\mu \alpha^\nu + \frac{1}{2} {\sigma_\mu}^\nu + \frac{1}{2} \theta {P_\mu}^\nu + \frac{1}{2} \omega {\varepsilon_\mu}^\nu .
\end{align}
Had we used the restricted connection such that $2 g_{\alpha [ \mu } \nabla_{\nu ] } v^\alpha = - F_{\mu \nu}$ not all this data would be independent, but $\alpha^\mu$ would be identified with the electric field and $\omega$ with the magnetic field.

Also note that we have not included a ``torsional magnetic field'' $G = \frac{1}{2} \varepsilon^{\mu \nu} G_{\mu \nu}$ in the list above since this is zero on causal backgrounds. One could certainly include this and compute away, but we do not particularly trust our results when there is no notion of time evolution. In particular the second law of thermodynamics would be essentially meaningless. In either case, the point is moot as including a nonzero $G$ does not introduce further constraints.

Finally, the most general first-order constitutive relations for the frame invariants consistent with spacetime symmetries are
\begin{align}
	&\mathcal S = - \zeta \theta - \tilde f_b b - \tilde f_B B - \tilde f_\omega \omega,\nonumber \\
	&\mathcal J^\mu = \sigma_e e^\mu + \sigma_E E^\mu + \sigma_G G^\mu  
	+ \sigma_T \nabla^\mu T + \sigma_Q \nabla^\mu \nu_Q + \sigma_M \nabla^\mu \nu_M ,\nonumber \\ 
	& \qquad \qquad +  \tilde \sigma_e  \tilde e^\mu +  \tilde \sigma_E \tilde  E^\mu +  \tilde \sigma_G \tilde  G^\mu  
	+  \tilde \sigma_T \tilde  \nabla^\mu T + \tilde  \sigma_Q \tilde  \nabla^\mu \nu_Q + \tilde  \sigma_M  \tilde \nabla^\mu \nu_M ,\nonumber \\
	&\mathcal E^\mu = \kappa_e e^\mu + \kappa_E E^\mu + \kappa_G G^\mu  
	+ \kappa_T \nabla^\mu T + \kappa_Q \nabla^\mu \nu_Q + \kappa_M \nabla^\mu \nu_M ,\nonumber \\ 
	& \qquad \qquad +  \tilde \kappa_e  \tilde e^\mu +  \tilde \kappa_E \tilde  E^\mu +  \tilde \kappa_G \tilde  G^\mu  
	+  \tilde \kappa_T \tilde  \nabla^\mu T + \tilde  \kappa_Q \tilde  \nabla^\mu \nu_Q + \tilde  \kappa_M  \tilde \nabla^\mu \nu_M ,\nonumber \\
	&\zeta^\mu = (\zeta_\theta \theta + \tilde \zeta_b b + \tilde \zeta_B B + \tilde \zeta_\omega \omega) u^\mu \nonumber \\
	&\qquad \qquad + \zeta_e e^\mu + \zeta_E E^\mu + \zeta_G G^\mu  
	+ \zeta_T \nabla^\mu T + \zeta_Q \nabla^\mu \nu_Q + \zeta_M \nabla^\mu \nu_M ,\nonumber \\ 
	& \qquad \qquad +  \tilde \zeta_e  \tilde e^\mu +  \tilde \zeta_E \tilde  E^\mu +  \tilde \zeta_G \tilde  G^\mu  
	+  \tilde \zeta_T \tilde  \nabla^\mu T + \tilde  \zeta_Q \tilde  \nabla^\mu \nu_Q + \tilde  \zeta_M  \tilde \nabla^\mu \nu_M ,\nonumber \\
	&\mathcal T^{\mu \nu} = - \eta \sigma^{\mu \nu} - \tilde \eta \tilde \sigma^{\mu \nu} .
\end{align}
In this we have defined the ``dual'' operation
\begin{align}
	\tilde v^\mu = \varepsilon^{\mu \nu} v_\nu ,
	&& \tilde w^{\mu \nu} = \varepsilon_\lambda{}^{(\mu } w^{\nu ) \lambda}
\end{align}
on vectors and symmetric two tensors. It has the properties
\begin{align}
	\tilde v_{1\mu} v_2^\mu = - v_{1 \mu} \tilde v_2^\mu ,
	&&\tilde w_{1\mu \nu} w_2^{\mu \nu} = - w_{1 \mu \nu} \tilde w_2^{\mu \nu}  .
\end{align}
We have similarly used tildes to label parity odd response coefficients, e.g.~$\sigma_E$ is the normal electrical conductivity and $\sigma_E$ is the Hall conductivity.

\subsection{Entropy Current Analysis}

Now let's move on to determining those constraints that result from imposing the second law of thermodynamics
\begin{align}
	( \nabla_\mu - G_\mu ) s^\mu \geq 0.
\end{align}
Start by considering the genuine second order data in the entropy production
\begin{align}
	(\nabla_\mu - G_\mu ) \zeta^\mu \big|_{2 - \partial} &= \zeta_\theta \dot \theta + \tilde \zeta_\omega \dot \omega + \tilde \zeta_b \dot b  + ( \tilde \zeta_B - \tilde \zeta_E ) \dot B + \tilde \zeta_e \varepsilon^{\mu \nu} \nabla_\mu e_\nu \nonumber \\
	&+ \zeta_T \nabla^2 T + \zeta_Q \nabla^2 \nu_Q + \zeta_M \nabla^2 \nu_M + \zeta_e \nabla_\mu e^\mu + \zeta_E \nabla_\mu E^\mu + \zeta_G \nabla_\mu G^\mu \geq 0 \nonumber
\end{align}
where we have used the identities
\begin{align}
	&\varepsilon^{\mu \nu} ( \nabla_\mu - G_\mu ) E_\nu = - \dot B - B \theta ,
	&&\varepsilon^{\mu \nu} \nabla_\mu  G_\nu = 0 .
\end{align}
The first is just Faraday's law and is equivalent to the closedness of $F$ (note we do not have a corresponding identity for $\ff$). The second similarly follows from $dG =0$ as well as $n \wedge dn = 0$.

We conclude that all the coefficients listed above must vanish and so $\zeta^\mu$ only has contributions from the remaining parity odd part
\begin{align}
	\zeta^\mu &= \tilde \zeta_B B u^\mu 
	+  \tilde \zeta_B \tilde  E^\mu +  \tilde \zeta_G \tilde  G^\mu  
	+  \tilde \zeta_T \tilde  \nabla^\mu T + \tilde  \zeta_Q \tilde  \nabla^\mu \nu_Q + \tilde  \zeta_M  \tilde \nabla^\mu \nu_M ,
\end{align}

The rest of the divergence of $\zeta^\mu$ is then
\begin{align}
	( \nabla_\mu - G_\mu ) \zeta^\mu = 
	&- \left( T \partial_\epsilon p \partial_T \tilde \zeta_B + \frac{1}{T} \partial_q p \partial_Q \tilde \zeta_B + \frac{1}{T} \partial_\rho p \partial_M \tilde \zeta_B \right) B \theta \nonumber \\
	&+ \partial_T \tilde \zeta_B \tilde E^\mu \nabla_\mu T + \partial_Q \tilde \zeta_B \tilde E^\mu \nabla_\mu \nu_Q + \partial_M \tilde \zeta_B \tilde E^\mu \nabla_\mu \nu_M \nonumber \\
	&+ ( \tilde \zeta_T + \partial_T \tilde \zeta_G ) \tilde G^\mu \nabla_\mu T + ( \tilde \zeta_Q + \partial_Q \tilde \zeta_G ) \tilde G^\mu \nabla_\mu \nu_Q + ( \tilde \zeta_M + \partial_M \tilde \zeta_G ) \tilde G^\mu \nabla_\mu \nu_M \nonumber \\
	&+ ( \partial_Q \tilde \zeta_T - \partial_T \tilde \zeta_Q ) \tilde \nabla^\mu T \nabla_\mu \nu_Q + ( \partial_M \tilde \zeta_T - \partial_T \tilde \zeta_M ) \tilde \nabla^\mu T \nabla_\mu \nu_M \nonumber \\
	&+ ( \partial_M \tilde \zeta_Q - \partial_Q \tilde \zeta_M ) \tilde \nabla^\mu \nu_Q \nabla_\mu \nu_M
\end{align}
which is supplemented by the canonical entropy production
\begin{align}
	( \nabla_\mu - G_\mu ) s^\mu_\text{can} &= \frac{1}{T} \zeta \theta^2 + \frac{1}{2 T} \eta \sigma_{\mu \nu} \sigma^{\mu \nu} \nonumber \\
	&+ \frac{1}{T} \sigma_E (E_\mu - T \nabla_\mu \nu_Q) (E^\mu - T \nabla^\mu \nu_Q ) - \frac{1}{T^2} \kappa_T ( \nabla_\mu T - T G^\mu ) ( \nabla^\mu T - T G^\mu ) \nonumber \\
	&+ \frac{1}{T} \tilde f_b b \theta + \frac{1}{T} \tilde f_B B \theta + \frac{1}{T} \tilde f_\omega \omega \theta 
	+ \frac{1}{T} \sigma_e e_\mu E^\mu  + \frac{1}{T} \kappa_e e^\mu G_\mu + \frac{1}{T} \left( \sigma_G + \kappa_E \right) E^\mu G_\mu  \nonumber \\
	&- \frac{1}{T^2} \kappa_e e^\mu \nabla_\mu T - \sigma_e e^\mu \nabla_\mu \nu_Q + \frac{1}{T} \left( \sigma_T - \frac{1}{T} \kappa_E \right) E^\mu \nabla_\mu T + \frac{1}{T} \sigma_M E^\mu \nabla_\mu \nu_M \nonumber \\
	&+ \left( \frac{1}{T} \kappa_Q - \sigma_G \right) G^\mu \nabla_\mu \nu_Q + \frac{1}{T} \kappa_M G^\mu \nabla_\mu \nu_M \nonumber \\
	&- \left( \sigma_T + \frac{1}{T^2} \kappa_Q \right) \nabla^\mu T \nabla_\mu \nu_Q - \frac{1}{T^2} \kappa_M \nabla^\mu T \nabla_\mu \nu_M - \sigma_M \nabla^\mu \nu_Q \nabla_\mu \nu_M  \nonumber \\
	&+ \frac{1}{T} \tilde \sigma_e \tilde e^\mu E_\mu + \frac{1}{T} \tilde \kappa_e \tilde e^\mu G_\mu - \frac{1}{T} ( \tilde \sigma_G - \tilde \kappa_E ) \tilde E^\mu G_\mu \nonumber \\
	&- \frac{1}{T^2} \tilde \kappa_e \tilde e^\mu \nabla_\mu T - \tilde \sigma_e \tilde e^\mu \nabla_\mu \nu_Q - \frac{1}{T} \left( \tilde \sigma_T + \frac{1}{T} \tilde \kappa_E \right) \tilde E^\mu \nabla_\mu T \nonumber \\
	 &- \left( \tilde \sigma_E + \frac{1}{T} \tilde \sigma_Q \right) \tilde E^\mu \nabla_\mu \nu_Q - \frac{1}{T} \tilde \sigma_M \tilde E^\mu \nabla_\mu \nu_M \nonumber \\
	 &- \frac{1}{T} \left( \tilde \kappa_T + \frac{1}{T} \tilde \kappa_G \right) \tilde G^\mu \nabla_\mu T - \left( \tilde \sigma_G + \frac{1}{T} \tilde \kappa_Q \right) \tilde G^\mu \nabla_\mu \nu_Q - \frac{1}{T} \tilde \kappa_M \tilde G^\mu \nabla_\mu \nu_M \nonumber \\
	 &- \left( \tilde \sigma_T - \frac{1}{T^2} \tilde \kappa_Q \right) \tilde \nabla^\mu T \nabla_\mu \nu_Q + \frac{1}{T^2} \tilde \kappa_M \tilde \nabla^\mu T \nabla_\mu \nu_M + \tilde \sigma_M \tilde \nabla^\mu \nu_Q \nabla_\mu \nu_M
\end{align}
In the above we have made the identifications $- \frac{1}{T} \sigma_Q = \sigma_E $ and $- \frac{1}{T} \kappa_G  = \kappa_T $ so that the entropy production due to electrical and thermal conductivity factors into a perfect square\footnote{In fact, what one should do is demand that the quadratic form defined by these transport coefficients be degenerate and positive semi-definite (degenerate so that that equilibrium solutions exist in non-zero background fields). This immediately gives these identities.}.

Demanding the second law then requires
\begin{align}
	&\zeta \geq 0 ,\qquad
	\eta \geq 0 ,\qquad
	\sigma_E \geq 0 ,\qquad
	\kappa_T \leq 0,  \qquad
	\tilde \sigma_G = \tilde \kappa_E ,\nonumber \\
	&\sigma_e = \sigma_M = \kappa_M = \kappa_e = \tilde \sigma_e = \tilde \kappa_e = 0 ,
	\qquad \qquad \kappa_E = - \sigma_G = - \frac{1}{T} \kappa_Q  = T \sigma_T ,\nonumber \\
	&
	\tilde f_\omega = \tilde f_b = 0 ,\qquad
	\tilde f_B = T^2 \partial_\epsilon p \partial_T \tilde \zeta_B + \partial_q p \partial_Q \tilde \zeta_B + \partial_\rho p \partial_M \tilde \zeta_B ,\nonumber \\
	%&\tilde f_\omega = T^2 \partial_\epsilon p \partial_T \tilde \zeta_\omega + \partial_q p \partial_Q \tilde \zeta_\omega + \partial_\rho p \partial_M \tilde \zeta_\omega \nonumber \\
	&\begin{pmatrix}
		\tilde \zeta_T + \partial_T \tilde \zeta_G \\
		\tilde \zeta_Q + \partial_Q \tilde \zeta_G \\
		\tilde \zeta_M + \partial_M \tilde \zeta_G
	\end{pmatrix} 
	=
	\begin{pmatrix}
		\frac{1}{T} \tilde \kappa_T + \frac{1}{T^2} \tilde \kappa_G \\
		\tilde \kappa_E + \frac{1}{T} \tilde \kappa_Q \\
		\frac{1}{T} \tilde \kappa_M
	\end{pmatrix},
	\qquad \qquad
	\begin{pmatrix}
		\partial_T \tilde \zeta_B \\
		\partial_Q \tilde \zeta_B \\
		\partial_M \tilde \zeta_B
	\end{pmatrix} 
	=
	\begin{pmatrix}
		\frac{1}{T} \tilde \sigma_T + \frac{1}{T^2} \tilde \kappa_E \\
		\tilde \sigma_E + \frac{1}{T} \tilde \sigma_Q \\
		\frac{1}{T} \tilde \sigma_M
	\end{pmatrix},
	\nonumber \\
	&\begin{pmatrix}
		\partial_Q \tilde \zeta_M - \partial_M \tilde \zeta_Q \\
		\partial_M \tilde \zeta_T - \partial_T \tilde \zeta_M \\
		\partial_T \tilde \zeta_Q - \partial_Q \tilde \zeta_T
	\end{pmatrix}
	=
	\begin{pmatrix}
		\tilde \sigma_M \\
		- \frac{1}{T^2} \tilde \kappa_M \\
		\frac{1}{T^2} \tilde \kappa_Q - \tilde \sigma_T
	\end{pmatrix}.
\end{align}

To untangle the differential constraints, begin by defining
\begin{align}\label{functions}
	&\tilde f = \tilde \zeta_B,
	&&T \tilde h_T = \tilde \zeta_T + \partial_T \tilde \zeta_G,
	&&T \tilde h_Q = \tilde \zeta_Q + \partial_Q \tilde \zeta_G + T \tilde f,
	&&T \tilde h_M = \tilde \zeta_M + \partial_M \tilde \zeta_G .
\end{align}
These then read
\begin{align}\label{pde}
	&\begin{pmatrix}
		\tilde h_T  \\
		\tilde h_Q  \\
		\tilde h_M
	\end{pmatrix} 
	=
	\begin{pmatrix}
		\frac{1}{T^2} \tilde \kappa_T + \frac{1}{T^3} \tilde \kappa_G \\
		\frac{1}{T} \tilde \kappa_E + \frac{1}{T^2} \tilde \kappa_Q + \tilde f\\
		\frac{1}{T^2} \tilde \kappa_M
	\end{pmatrix},
	\qquad \qquad
	\begin{pmatrix}
		\partial_T \tilde f \\
		\partial_Q \tilde f \\
		\partial_M \tilde f
	\end{pmatrix} 
	=
	\begin{pmatrix}
		\frac{1}{T} \tilde \sigma_T + \frac{1}{T^2} \tilde \kappa_E \\
		\tilde \sigma_E + \frac{1}{T} \tilde \sigma_Q \\
		\frac{1}{T} \tilde \sigma_M
	\end{pmatrix},
	\nonumber \\
	&\begin{pmatrix}
		\partial_Q \tilde h_M - \partial_M \tilde h_Q \\
		\partial_M \tilde h_T - \partial_T \tilde h_M \\
		\partial_T \tilde h_Q - \partial_Q \tilde h_T
	\end{pmatrix}
	=
	\begin{pmatrix}
		\frac{1}{T} \tilde \sigma_M - \partial_M \tilde f\\
		\frac{1}{T} \tilde h_M - \frac{1}{T^3} \tilde \kappa_M \\
		- \frac{1}{T} \tilde h_Q + \frac{1}{T^3} \tilde \kappa_Q - \frac{1}{T}\tilde \sigma_T + \partial_T \tilde f + \frac{1}{T} \tilde f
	\end{pmatrix} .
\end{align}
This leads to several consistency relations on the four functions (\ref{functions})
\begin{align}\label{consistency}
	&\begin{pmatrix}
		\partial_Q \tilde h_M - \partial_M \tilde h_Q \\
		\partial_M \tilde h_T - \partial_T \tilde h_M \\
		\partial_T \tilde h_Q - \partial_Q \tilde h_T
	\end{pmatrix}
	=
	\begin{pmatrix}
		0\\
		0 \\
		0
	\end{pmatrix} .
\end{align}
The first comes from comparing the final component of the second equation to the first component of the third while the second follows from comparing the final component of the first equation to the second of the third. The final condition results from combining the second component of the first equation, the first component of the second and the final component of the third.
The vector $\begin{pmatrix} \tilde h_T & \tilde h_Q & \tilde h_M \end{pmatrix}^T$ is then curl free as so must be the gradient of some function $\tilde g ( T , \nu_Q , \nu_M )$
\begin{align}
	\tilde h_T = \partial_T \tilde g,
	&&\tilde h_Q = \partial_Q \tilde g ,
	&&\tilde h_M = \partial_M \tilde g .
\end{align}

\subsection{Summary of Results}\label{subsec:fluid_summary_multi}
This solves the full set of restrictions imposed by the second law. Before summarizing results, the following redefinition of transport coefficients will simplify the final answer
\begin{align}
	&T \tilde f \rightarrow \tilde m,
	&&T^2 \tilde g \rightarrow \tilde m_\epsilon , \nonumber \\
	&\tilde \sigma_ T \rightarrow \tilde \sigma_T + \partial_T \tilde m,
	&&\tilde \kappa_T \rightarrow \tilde \kappa_T + \partial_T \tilde m_\epsilon . \nonumber \\
\end{align}
after which frame invariants are 
\begin{align}\label{frame invariants summary}
	&\mathcal T^{\mu \nu} = - \eta \sigma^{\mu \nu} - \tilde \eta \tilde \sigma^{\mu \nu} \qquad \qquad \qquad \qquad
	\mathcal S = - \zeta \Theta - \tilde f_B B \nonumber \\
	&\mathcal J^\mu = \sigma_E \left( E^\mu - T \nabla^\mu \nu_Q \right) + \sigma_T ( \nabla^\mu T - T G^\mu ) +  \tilde \sigma_E \left( \tilde  E^\mu - T \tilde \nabla^\mu   \nu_Q \right) + \tilde  \sigma_T ( \tilde  \nabla^\mu T - T \tilde  G^\mu ) \nonumber \\
	& \qquad \qquad   - \tilde m  \tilde G^\mu + \tilde \nabla^\mu \tilde m \nonumber \\
	&\mathcal E^\mu = T \sigma_T \left( E^\mu - T \nabla^\mu \nu_Q \right) +  \kappa_T  ( \nabla^\mu T - T G^\mu )   -T \tilde \sigma_T \left( \tilde E^\mu - T \tilde \nabla^\mu \nu_Q \right) +  \tilde \kappa_T  ( \tilde \nabla^\mu T - T \tilde G^\mu )     \nonumber \\
	&\qquad \qquad  
	- \tilde m \tilde E^\mu - 2 \tilde m_\epsilon \tilde G^\mu  + \tilde \nabla^\mu \tilde m_\epsilon .
\end{align}

The most general set of first order transport coefficients are then as follows. There are four sign semi-definite functions of all three thermodynamic variables
\begin{align}
	\zeta \geq 0,
	&&\eta \geq 0,
	&&\sigma_E \geq 0 ,
	&&\kappa_T \leq 0.
\end{align}
These are the bulk viscosity, shear viscosity, conductivity and thermal conductivity, all of which are zero for dissipationless fluids.
One sign indefinite parity-even coefficient exists, a thermo-electric coefficient
\begin{align}
	\sigma_T 
\end{align}
which determines the charge flow due to thermal gradients and the energy flow due to electromagnetic fields.

The parity odd sector is richer, including six unconstrained parity odd coefficients
\begin{align}
	\tilde \eta,
	&&\tilde \sigma_E,
	&&\tilde \kappa_T,
	&&\tilde \sigma_T , 
	&& \tilde m,
	&& \tilde m_\epsilon,
\end{align}
a Hall viscosity, Hall conductivity, thermal Hall conductivity, thermo-electric Hall coefficient, magnetization and energy magnetization. $\tilde m$ and the equation of state determine the magnetic field induced pressure
\begin{align}
	&\tilde f_B =T^2 \partial_\epsilon p  \partial_T \left(  \frac{\tilde m}{T} \right) + \partial_q p \partial_Q  \left(  \frac{\tilde m}{T} \right) + \partial_\rho p \partial_M  \left(  \frac{\tilde m}{T} \right) .
\end{align}

The attentive reader may note that the parity odd response to thermal gradients differs from the parity odd response to the Luttinger potential by the energy magnetization (and similarly for the thermoelectric Hall coefficient), as compared to the well-known results of \cite{Kubo:1957yq,Luttinger:1964zz}. This is because these works assumed vanishing equilibrium currents, as pointed out in the footnote below equation (4.10) of \cite{Kubo:1957yq}. In general, pure curl persistent equilibrium currents may arise, given by the magnetizations, and in this case the proper relationship is that given above \eqref{frame invariants summary}.

To get a feel for these results, it's helpful to fix a fluid frame and write the results for the non-covariant currents defined in (\ref{stress-energy}). We choose our frame so the physical mass, charge and energy correspond with the thermodynamic ones and the velocity is that of the mass current
\begin{align}\label{frame}
	\mathcal Q = \varrho =  \mathcal E = 0, && \mu^\mu = 0 .
\end{align} 
The frame invariants $\mathcal J^\mu$ and $\mathcal E^\mu$ are then simply the first-order deviations $\nu^\mu $ and $\xi^\mu$. Using (\ref{nc currents}) to retrieve the non-covariant currents from this data we have
\begin{align}
	&\rho^0 = \rho, \qquad \qquad  \rho^i = \rho u^i
	\qquad \qquad j^0 = q , 
	\qquad \qquad  \varepsilon_\nc^0 = \frac{1}{2} \rho u^2 + \epsilon ,\nonumber \\
	 &j^i = q u^i + \sigma_E (e^\Phi  E^i + B \varepsilon^{ij} u_j - T \partial^i \nu_Q ) + \tilde \sigma_E \varepsilon^{ij} ( e^\Phi E_j + B \varepsilon_{jk} u^k - T \partial_j \nu_Q ) \nonumber \\
	 &\qquad+ \sigma_T e^\Phi \partial^i ( e^{-\Phi} T ) + \tilde \sigma_T e^\Phi \varepsilon^{ij} \partial_j (e^{-\Phi} T) + e^\Phi \varepsilon^{ij} ( e^{-\Phi} \tilde m ) , \nonumber \\
	 &\varepsilon_\nc^i = \left( \frac{1}{2} \rho u^2 + \epsilon + p - \zeta \theta - \tilde f_B B \right) u^i - \eta \sigma^{ij} u_j - \tilde \eta \tilde \sigma^{ij} u_j  \nonumber \\
	 &\qquad + T \sigma_T ( e^\Phi  E^i + B \varepsilon^{ij} u_j - T \partial^i \nu_Q ) -T \tilde \sigma_T \varepsilon^{ij} ( e^\Phi E_j + B \varepsilon_{jk} u^k - T \partial_j \nu_Q ) \nonumber \\
	 &\qquad   + \kappa_T e^\Phi \partial^i ( e^{-\Phi} T ) + \tilde \kappa_T e^\Phi \varepsilon^{ij} \partial_j ( e^{- \Phi} T ) 
	  - \tilde m \varepsilon^{ij} ( e^\Phi E_j + B \varepsilon_{jk} u^k ) + e^{2 \Phi} \varepsilon^{ij} \partial_j ( e^{-2 \Phi } \tilde m_\epsilon ) , \nonumber \\
	 & T_\nc^{ij} = \rho u^i u^j + ( p - \zeta \theta - \tilde f_B B  ) h^{ij} - \eta \sigma^{ij} - \tilde \eta \tilde \sigma^{ij} .\label{eq:non-cov_currents_multi}
\end{align}

In \eqref{eq:non-cov_currents_multi} above and in what follows $E^i$ is defined to be the electric field in the lab frame so that the comoving electric field used earlier is\footnote{We hope the careful reader will forgive the notational dissonance.}
\begin{align}
	\begin{pmatrix}
		0 & - E_j \\
		E_i & B \varepsilon_{ij} 
	\end{pmatrix}
	\begin{pmatrix}
		e^\Phi \\
		u^j
	\end{pmatrix}
	=
	\begin{pmatrix}
		- E_j u^j \\
		e^\Phi E_i + B \varepsilon_{ij} u^j
	\end{pmatrix} .
\end{align}
$\theta$ and $\sigma^{ij}$ are the curved space quantities defined in (\ref{decomposition}) .
Using the connection (\ref{christoffels}) we find that 
\begin{align}
	\nabla^i u^j &= \partial^i u^j + \Gamma^{ji}{}_\lambda u^\lambda = \partial^i u^j + \Gamma^{ji}{}_k u^k + e^\Phi \Gamma^{ji}{}_0 , \nonumber \\
	&= \hat\nabla^i u^j - u^j \hat\nabla^i \Phi - \frac{1}{2} \dot h^{ij} + e^\Phi \hat\nabla^{[j}( e^{- \Phi} u^{i]} )+ \frac{1}{2} \Omega^{ ij} , \nonumber \\
	&= e^\Phi \hat\nabla^{(i} ( e^{- \Phi} u^{j)} ) - \frac{1}{2}  \dot h^{ij} + \frac{1}{2} \Omega^{ ij} , \nonumber \\
	\implies \qquad  \sigma^{ij} &= e^\Phi \hat\nabla^i ( e^{-\Phi} u^j ) + e^\Phi \hat\nabla^j ( e^{-\Phi} u^i ) - \dot h^{ij}  - h^{ij} \theta
\end{align}
where $\hat\nabla_{i}$ is the covariant derivative on a spatial slice and
\begin{align}
	&\theta = \frac{1}{\sqrt{h} e^{- \Phi}} \partial_\mu ( \sqrt{h} e^{-\Phi} u^\mu) = e^\Phi  \hat\nabla_i ( e^{- \Phi} u^i ) + \frac{1}{2} e^\Phi h^{ij} \dot h_{ij} .
\end{align}

\subsection{Results for a Single Component Fluid}\label{sec:single component}
A single component fluid satisfies additional constraints since the charge and mass currents are proportional and we investigate these constraints in this section. Since this is also the case of applicability for \cite{Jensen:2014ama} it will allow for a direct comparison of our results.

Let the single constituent be of charge $e$ and mass $m$. The charge density and mass density are then related to a single function $n$, the number density
\begin{align}
	q = e n ,
	&&\rho = m n
\end{align}
and the thermodynamic relation (\ref{thermo2}) takes the form
\begin{align}
	d p  = \frac{\epsilon + p}{T} d T + T n d \nu
\end{align}
where $\nu = e \nu_Q + m \nu_M $ is the total chemical potential. All thermodynamic functions must be a function of this combination.

Now since $ m \nu^\mu = e \mu^\mu$, the vector frame invariant $\mathcal J^\mu$ is zero which gives the restrictions
\begin{align}
	\sigma_E = \sigma_T = \tilde \sigma_E = 0 ,
	&& \tilde m = \tilde m (T ) , 
	&& \tilde \sigma_T = - \tilde m'  ,
	&&T \tilde \sigma_T = - \tilde m  .
\end{align}
The latter two relations imply that $\tilde m$ is a linear function of $T$ and so that $\tilde \sigma_T$ is a constant independent of the thermodynamic state variables.

This simplifies $\mathcal E^\mu$ to
\begin{align}
	&\mathcal E^\mu =  \kappa_T  ( \nabla^\mu T - T G^\mu )  +  \tilde \kappa_T  ( \tilde \nabla^\mu T - T \tilde G^\mu ) + T^2 \tilde \sigma_T \tilde \nabla^\mu \nu_Q 
	 - 2 \tilde m_\epsilon  \tilde G^\mu  + \tilde \nabla^\mu \tilde m_\epsilon .
\end{align}
Since we cannot have dependence on $\nu_Q$ outside of the combination $\nu$, we must also have $\tilde \sigma_T = 0$.

In summary, for a single component fluid there are three sign semi-definite functions including a thermal conductivity and two viscosities
\begin{align}
	\zeta \geq 0 , &&
	\eta \geq 0 , && 
	\kappa_T \leq 0 
\end{align}
which exhausts the parity even sector. The parity odd sector contains three free functions including a Hall viscosity, thermal Hall conductivity and energy magnetization
\begin{align}
	\tilde \eta,
	&&\tilde \kappa_T,
	%&&\tilde \zeta_\omega , 
	&& \tilde m_\epsilon.
\end{align}
In terms of these coefficients, the frame invariants are
\begin{align}
	& \mathcal S = - \zeta \theta \qquad \qquad \qquad \qquad
	\mathcal T^{\mu \nu} = - \eta \sigma^{\mu \nu} - \tilde \eta \tilde \sigma^{\mu \nu}\nonumber \\
	&\mathcal E^\mu =  \kappa_T  ( \nabla^\mu T - T G^\mu )  +  \tilde \kappa_T  ( \tilde \nabla^\mu T - T \tilde G^\mu ) - 2 \tilde m_\epsilon  \tilde G^\mu  + \tilde \nabla^\mu \tilde m_\epsilon .
\end{align}
One may equivalently state this result in terms of a conductivity $\sigma_E$ and a Hall conductivity $\tilde \sigma_E$ by exchanging the first order data $\nabla^\mu T - T G^\mu$ for $e E^\mu - T \nabla^\mu \nu - m \alpha^\mu$ using the equations of motion (\ref{Navier}).

Finally, the noncovariant currents in the frame (\ref{frame}) are
\begin{align}
	&j^0 = q,\qquad \qquad j^i = q u^i ,
	\qquad \qquad  \varepsilon_\nc^0 = \frac{1}{2} \rho u^2 + \epsilon  ,\nonumber \\
	 & \varepsilon_\nc^i = \left( \frac{1}{2} \rho u^2 + \epsilon + p - \zeta \theta  \right) u^i - \eta \sigma^{ij} u_j - \tilde \eta \tilde \sigma^{ij} u_j  \nonumber \\
	 &\qquad  + \kappa_T e^\Phi \partial^i ( e^{-\Phi} T ) + \tilde \kappa_T e^\Phi \varepsilon^{ij} \partial_j ( e^{- \Phi} T ) 
	   + e^{2 \Phi} \varepsilon^{ij} \partial_j ( e^{-2 \Phi } \tilde m_\epsilon ) , \nonumber \\
	 & T_\nc^{ij} = \rho u^i u^j + ( p - \zeta \theta ) h^{ij} - \eta \sigma^{ij} - \tilde \eta \tilde \sigma^{ij} .
\end{align}
This differs from the results (1.13-18) of \cite{Jensen:2014ama}, which includes two parity odd parameters besides the Hall viscosity and thermal Hall conductivity and a differential relationship determining the magnetic field induced pressure, a transverse $E_i$ term and a curl $\partial_i T$ in terms of them.

We have also checked the results for lowest Landau level fluids \cite{Geracie:2014zha} with the alterations to the derivative operator discussed here and found that they survive without change. This can be understood in part in terms of our equations of motion, which differ from \cite{Jensen:2014aia} as the acceleration $\alpha^\mu$ is independent data and not tied to $E^\mu$. Projection to the lowest Landau level (which may be thought of as a massless limit) removes the acceleration terms in Navier-Stokes equation (\ref{Navier}) and we reproduce the constraint equation
\begin{align}
	\nabla^\mu p =  q E^\mu + ( \epsilon + p ) G^\mu 
\end{align}
used in \cite{Geracie:2014zha}. The only change to the calculation would then be in the available first order data, which is augmented in our approach. However, a detailed calculation does show that the new terms drop out after the entropy current analysis and we retrieve the previous results.

\subsection{Kubo formulas}
The transport coefficients we have found are quite familiar and have been subjected to extensive study in the literature and calculated for a number of systems. Calculation from a microscopic theory usually proceeds by the use of Kubo formulas. The techniques to derive these are now standard and Kubo formulas for all the transport coefficients presented above have been given in the literature. Here we present them in our notation and in the frame \eqref{frame} for the readers convenience.

The relevant retarted correlators, including contact terms are
\begin{align}
	G^{i j, k l} (x) &= \left \langle \frac{\delta T^{ij} (x) }{ \delta g_{kl} (0) }\right \rangle+ \frac 1 2 i \theta ( x^0 ) \left \langle \left[ T^{ij} (x) , T^{kl} (0) \right] \right \rangle ,\nonumber \\
	G^{\mu, \nu}_{jj} (x) &= \left \langle \frac{\delta j^\mu (x)}{\delta A_\nu (0)} \right \rangle + i \theta ( x^0 ) \left \langle \left[ j^\mu (x) , j^\nu (0) \right] \right \rangle , \nonumber \\
	G^{\mu ,\nu}_{j\varepsilon}(x) &= \left \langle \frac{\delta j^\mu (x)}{\delta n_\nu (0)} \right \rangle - i \theta ( x^0 ) \left \langle \left[ j^\mu (x) , \varepsilon^\nu (0) \right] \right \rangle , \nonumber \\
	G^{\mu, \nu}_{\varepsilon \varepsilon}(x) &= \left \langle \frac{\delta \varepsilon^\mu (x)}{\delta n_\nu (0)} \right \rangle - i \theta ( x^0 ) \left \langle \left[ \varepsilon^\mu (x) , \varepsilon^\nu (0) \right] \right \rangle .
\end{align}

In terms of these, the viscosities are
\begin{align}
	 &\zeta = - \lim_{\omega \rightarrow 0} \frac{\delta_{ij} \delta_{kl} G^{ij,kl} (\omega_+)}{2 i \omega_+}, 
	 &\eta = - \lim_{\omega \rightarrow 0} \frac{ \Pi_{ijkl} G^{ij,kl} (\omega_+)}{2 i \omega_+}, 
	 &&\tilde \eta = - \lim_{\omega \rightarrow 0} \frac{ \tilde \Pi_{ijkl} G^{ij,kl} (\omega_+)}{2 i \omega_+},
\end{align}
where we have introduced the projectors
\begin{align}
	\Pi^{ijkl} = \delta^{i(k} \delta^{l)j} - \frac 1 2 \delta^{ij} \delta^{kl} , 
	&&\tilde \Pi^{ijkl} = \frac 1 2 \left( \delta^{i ( k } \epsilon^{l) j} +\delta^{j(k} \epsilon^{l)i} \right)
\end{align}
and $\omega_+ = \omega + i \delta$ for a small, positive $\delta$. We recommend \cite{Bradlyn:2012ea} for a careful computation of these Kubo formulas.

The equations for the conductivity and thermoelectric conductivities are prototypical examples and first found in \cite{Kubo:1957uq,Kubo:1957yq}
\begin{align}
	&\sigma_E = \lim_{\omega \rightarrow 0} \frac{\delta_{ij} G^{i,j}_{jj} ( \omega_+ )}{2 i \omega_+} ,
	&& \tilde \sigma_E = \lim_{\omega \rightarrow 0} \frac{\epsilon_{ij} G^{i,j}_{jj} ( \omega_+ )}{2 i \omega_+} , \nonumber \\
	&T \sigma_T = \lim_{\omega \rightarrow 0} \frac{\delta_{ij} G^{i,j}_{j\varepsilon} ( \omega_+ )}{2 i \omega_+} ,
	&& T \tilde \sigma_T + \tilde m = \lim_{\omega \rightarrow 0} \frac{\epsilon_{ij} G^{i,j}_{j\varepsilon} ( \omega_+ )}{2 i \omega_+} .
\end{align}
Kubo formulas for the thermal conductivities were first computed in the classic work \cite{Luttinger:1964zz} where the Luttinger potential was introduced. We find them to be
\begin{align}
	&T \kappa_T = \lim_{\omega \rightarrow 0} \frac{\delta_{ij} G^{i,j}_{\varepsilon \varepsilon} ( \omega_+ )}{2 i \omega_+} ,
	&& T \tilde \kappa_T + 2 \tilde m_\epsilon = \lim_{\omega \rightarrow 0} \frac{\epsilon_{ij} G^{i,j}_{\varepsilon \varepsilon} ( \omega_+ )}{2 i \omega_+} .
\end{align}
As discussed in section \ref{subsec:fluid_summary_multi} the themoelectric and thermal Hall conductivities differ from the parity odd response to the chemical and Luttinger potentials by magnetizations $\tilde m$ and $\tilde m_\epsilon$ respectively, unlike the standard formulas found in \cite{Kubo:1957yq,Luttinger:1964zz}. The  Kubo formulas for $\tilde \sigma_T$ and $\tilde \kappa_T$ are therefore completed by expressions for the magnetizations, derived in \cite{Geracie:2014zha}:
\begin{align}
	& \tilde m - T \partial_T \tilde m = - \lim_{| \mathbf k | \rightarrow 0} \frac{i \epsilon_{ij} k^i G^{j,0}_{j \varepsilon} ( \mathbf k )}{|\mathbf k |^2} ,
	&& 2 \tilde m_\epsilon - T \partial_T \tilde m_\epsilon = - \lim_{| \mathbf k | \rightarrow 0} \frac{i \epsilon_{ij} k^i G^{j,0}_{\varepsilon \varepsilon} ( \mathbf k )}{|\mathbf k |^2} .
\end{align}

\section{Outlook}

In this paper we have considered the most general geometric backgrounds consistent with local Galilean invariance and developed the theory of first order dissipative fluids on such a background. This formalism at hand, there are a number of prospects for further investigation. One direct application would be to perform the fluid analysis carried out here for more general systems. Non-relativistic superfluids and superfluid/normal fluid mixtures as arise for instance in partially condensed superfluid Helium and it would be interesting to see the restrictions imposed by Galilean symmetry.

We have also presented a program for writing down invariant actions of massive fields that realize the Galilean group linearly. These are not however the most general actions consistent with non-relativistic symmetries. For instance, when matter is charged under boosts $\psi \rightarrow e^{i k^a K_a }\psi$ one may write down a non-relativistic form of the Dirac equation that is linear in both time and space derivatives \cite{Niederle:2007xp}.

One might also consider systems with spontaneously broken symmetries in which the Galilean group is realized non-linearly on a collection of Goldstones. It's long been understood how to write down the most general actions for a non-relativistic Goldstones to lowest order in derivatives \cite{Greiter:1989qb} but we believe use of the extended representation will prove useful in constructing actions to any order.

It would also be instructive to consider the most general effective actions one may write in terms of the background fields $(e^A, \omega^A{}_B , \aa , A )$.  Consider for example the (universal sector of the) effective field theory of the fractional quantum Hall effect \cite{Wen:1992ej,Son:2013,Abanov:2014ula},
\begin{align}
	S = \frac{1}{4 \pi}\int \left( \nu A \wedge d A + \kappa \omega \wedge d A + \kappa' \omega \wedge d \omega \right)
\end{align}
where $\omega = \frac 1 2 \varepsilon^{ab} \omega_{ab}$. This is a perfectly sensible effective action for the background fields $(A , e^a)$ and useful for studying electric and viscous transport.

However, studying massive transport also requires coupling to the $U(1)_M$ gauge field $\aa$. Since the microscopic action for a minimally coupled single component system always contains $A$ and $\aa$ in the combination $A + \frac{m}{q} \aa$ (see equation (\ref{cov div})) one might expect that the correct effective action contains only this function of $A$ and $\aa$. The resulting action is however not boost invariant and so physically unacceptable. Upon identifying boost transformations with the anomalous diffeomorphisms of \cite{Son:2005rv}, this problem is essentially the one considered in \cite{Hoyos:2011ez} where the problem is solve by improving the action order by order in derivatives so as to impose invariance. It would be useful to have a manifestly geometrical way to write this term and so generate these corrections using the technology developed here.

\section*{Acknowledgments}
It is a pleasure to thank Kristan Jensen, Dam T. Son and Robert M. Wald  for helpful discussions. M.G. is supported in part by NSF grant DMR-MRSEC 1420709. K.P. is supported in part by NSF grant PHY 12-02718. M.M.R is supported in part by DOE grant DE-FG02-13ER41958.

\bibliographystyle{JHEP}
\bibliography{WenZeerefs}

%%%%%%%%%%%%%%%%%%%%%%%%%%%%%%%%%%%%%%%%%%%%%%%%%%%%%%%%%%%%%%%%%
\end{document}